# The hidden spin-momentum locking and topological defects in unpolarized light fields

Peng Shi[1,†,*], Min Lin[1,†], Xinxin Gou[1,†], Luping Du[1,*], Aiping Yang[2], and Xiaocong Yuan[1,3,*]

[1]*Nanophotonics Research Centre, Institute of Microscale Optoelectronics & State Key Laboratory of Radio Frequency Heterogeneous Integration, Shenzhen University, Shenzhen 518060, China*

[2]*Research Institute of Interdisciplinary Science & School of Materials Science and Engineering, Dongguan University of Technology, Dongguan 523808, China*

[3]*Zhejiang Lab, Research Center for Humanoid Sensing, Research Institute of Intelligent Sensing, Hangzhou 311100, China*

Electromagnetic waves characterized by intensity, phase, and polarization degrees of freedom are widely applied in data storage, encryption, and communications. However, these properties can be substantially affected by disorders and disturbances, whereas high-dimensional structured light constructed by momentum and angular momentum degrees of freedom can offer new insights into the features and phenomena of electromagnetic waves, for example topological characteristics and structures that are robust to these disturbances. Here, we discover and demonstrate theoretically and experimentally the hidden spin–momentum locking and topological defects in unpolarized monochromatic light for the first time. The coherent spin angular momentum density is locked to the kinetic momentum density except for a small coupling spin term, due to the simultaneous presence of transverse magnetic and electric components. To cancel the coupling term, we employ a metal film acting as a polarizer to form the skyrmion-like spin textures at the metal/air interface. Using an in-house scanning microscopic system to image the out-of-plane coherent spin density of the unpolarized light, we obtained experimental results that coincide with our theoretical predictions. The theory and technique promote the applications of topological defects in anti-disturbance optical data storage, encryption, and free-space communications.

## I. INTRODUCTION

Momentum and angular momentum (AM), which are two fundamental dynamical properties of the electromagnetic (EM) field, play an important role in understanding and predicting light–matter interactions [1–4]. Generally, angular momentum can be decomposed into spin angular momentum (SAM) associated with circular polarization and orbital angular momentum (OAM) carried by the vortex phase [5]. The pairwise coupling of momentum, SAM, and OAM has resulted in a plenitude of remarkable phenomena, including the spin Hall effect of light [6,7], the optical Magnus effect [8], spin–orbit AM conversions [9,10], the quantum spin Hall effect of light [11–14], and optical topological defects [15–21]. It has led to the emergence of interesting research fields of optical spin-–orbit interactions and offers potential applications in manipulation [22–24], imaging [25,26], detection [27], metrology [28], data storage [29], encryption [30–35], and communications [36–38].

Previously, most of the research on optical spin–orbit interactions rely on pure polarized light, whereas the spin–orbit properties in a random or unpolarized field are rarely involved [39–42]. However, unpolarized fields are prevalent in optical data encoding, encryption, and free-space communications [30–38]. For example, consider encrypted structured light obtained by passing it through a random phase plate (RPP) with different Fresnel transmission coefficients and phase delays for the transverse electric (TE) and transverse magnetic (TM) polarization components respectively. The information encoded in the incident light by the intensity, phase, and polarization degrees of freedom is hard to be decrypted without auxiliary algorithms and keys because the transmission light is completely chaotic [35]. Nevertheless, the high-dimensional information in structured light [43], including the AM properties and the topological structure, can be protected even in the random EM system [44–46], which allows robust encryption and decryption in optical encryption systems and is also beneficial in applications in free-space optical communications.

In this work, we first demonstrate spin–momentum locking of unpolarized light, i.e., its coherent SAM density is locked with the Abraham–Poynting coherent kinetic momentum (KM) density even for focused unpolarized light, up to a small coupling term determined by the product of Fresnel transmission coefficients for the TM and TE-polarization components. This locking property satisfies the right-hand rule. Curiously, in contradiction with earlier results [39], an extraordinary axial SAM of unpolarized vortex light generated by passing linear polarized vortex light through a RPP establishes itself as the numerical aperture (NA) of objective lens increases. This makes it possible to form topological structures with unpolarized light. By utilizing a metal film to break the coupling between the TM- and TE-polarization components, various topological spin textures form at the metal/air interface. However, these topological spin textures are different from those produced by pure polarized light. We developed an in-house near-field optical scanning microscopic system to image the out-of-plane SAM densities of these topological spin textures. The experimental results match well with those from theory. The methods and techniques proposed here can be applied to the fields of robust optical data encoding, encryption, and communications based on topological defects.

## II. FORMATION OF PHOTONIC SKYRMION

To begin, we consider monochromatic linear polarized light passing through a RPP for which the amplitude transmission coefficients and phase delays for TE and the TM polarized components are completely random [see **FIG. 1**(a)]. By calculating the degree of polarization $\mathbb{P}$ and the complex degree of coherence $\mathscr{M}_{12}$ of the transmission light at the plane H, we find that $\mathbb{P} = 0$ and $\mathscr{M}_{12} = 0$ simultaneously [47, SM. I]. Thus, for the transmission light, its amplitude and phase are completely disordered, and the degree of polarization and the spatial coherence also vanishes. Therefore, this transmission light can be considered as unpolarized light.

Nevertheless, the SAM density and KM density can be resolved because of the self-correlation property of this unpolarized light [47, SM. II]. Here, we investigate the spin–momentum properties of this unpolarized light focused using an objective lens. After some lengthy calculations, we reached the conclusion that the coherent SAM density and KM density of this unpolarized light satisfy the relation [47, SM. III]:

$$\mathbf{S} = \frac{1}{2k^2}\left(\nabla \times \mathbf{P}\right) + \mathbf{S}_c . \qquad (1)$$

Here, $\mathbf{S} = \mathrm{Im}\{\varepsilon\mathbf{E}^* \times \mathbf{E} + \mu\mathbf{H}^* \times \mathbf{H}\}/4\omega$ is the SAM density; the KM density $\mathbf{P} = \mathrm{Re}\{\mathbf{E}^* \times \mathbf{H}\}/2v^2$ with $\mathbf{E}/\mathbf{H}$ the electric/magnetic field, $\varepsilon/\mu$ the permittivity/permeability of medium, $\omega$ the angular frequency, and $v$ the velocity of light in medium, respectively. The symbol Re/Im signifies the real/imaginary part of a complex number. This expression reveals that, even for unpolarized light, the spin is locked with the KM except for a very small term $\mathbf{S}_c$ [**FIG. 1**(b–d)]. This term $\mathbf{S}_c$ represents the coupling between the TE and the TM polarized components because it is related to the product of the Fresnel transmission coefficients.

Moreover, it is also determined by the numerical aperture (NA) of the objective lens. Therefore, to cancel or reduce this coupling term, one utilizes a polarizer to break the electric-magnetic dual symmetry or an objective lens with a small NA [47, FIG. S2]. In addition, Equation (1) suggests several methods to manipulate the SAM of unpolarized light: 1. The propagation direction can be controlled by adding an additional propagating phase to the incident light (because the spin is locked with the KM, the spin is inverted or inclined if the KM is reversed or tilted); and 2. The distribution of the KM density can be achieved by altering the EM modes or employing the multi-wave interference method (because the SAM density is proportional to the transverse gradient of the KM, the SAM density can be inverted or inclined by tuning the inhomogeneities of the KM density).

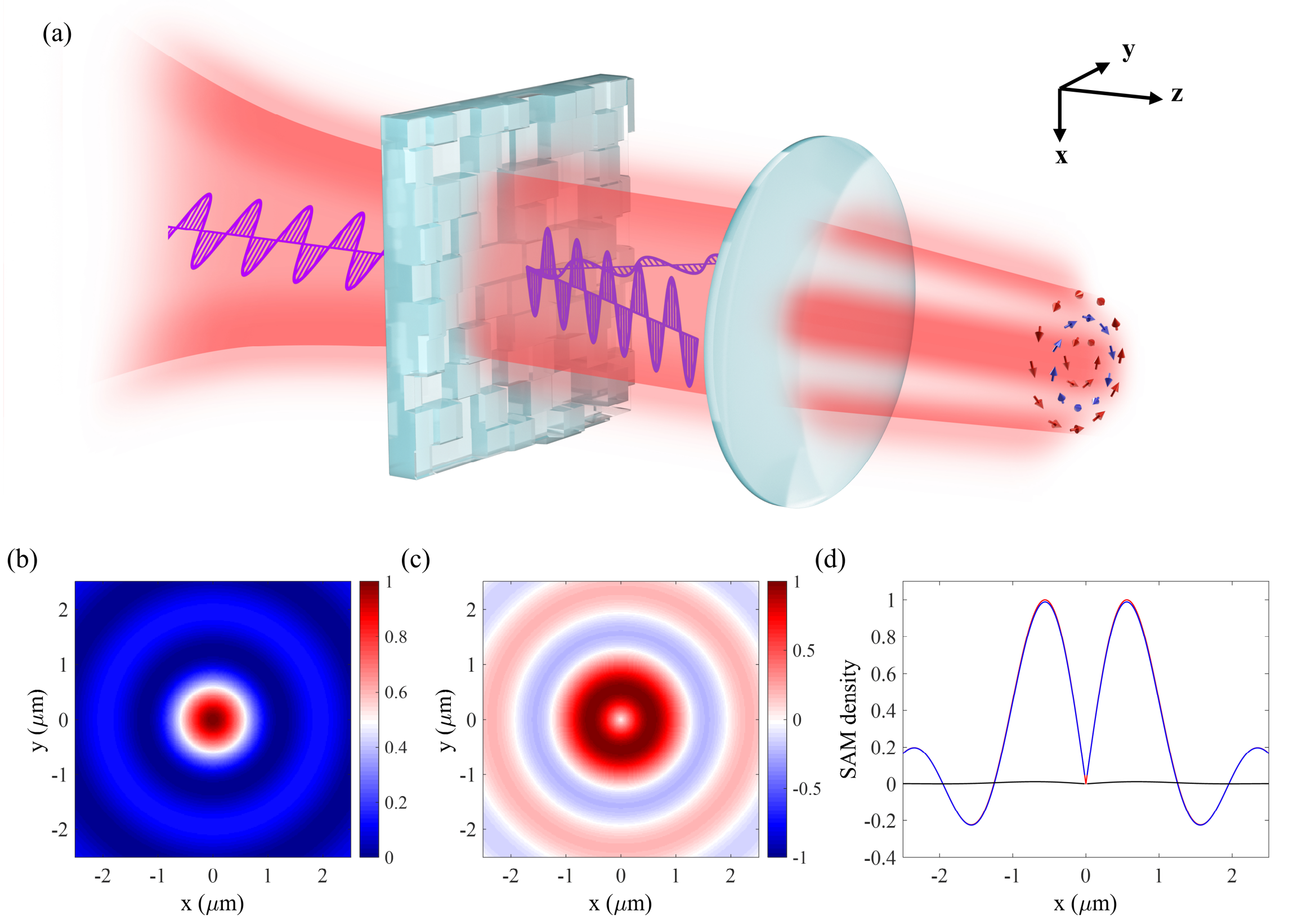

**FIG. 1.** Spin–momentum locking of unpolarized light. (a) Schematic of the generation of unpolarized light by passing linear polarized light through a RPP, which is subsequently focused by an objective lens; (b, c) density plots of the coherent KM and the coherent SAM density obtained with an objective lens of NA = 0.2, and (d) the one-dimensional (1D) contour plot of the total SAM density (red line), SAM density calculated using $\nabla\times\mathbf{P}/2k^2$ (blue line), and the coupling SAM density $\mathbf{S}_c$ (black line) indicate that, even for unpolarized light, spin is locked with KM except for a very small coupling term.

The optical vortex is widely applied in encoding and encryption applications [30–35]. Here, we also add a vortex phase with topological charge $\ell$ to the incident linear-polarized light [**FIG. 1**(a)]. In this instance, spin–momentum locking given by Equation (1) is also fulfilled. Besides the axial KM density [**FIG. 2**(a)], the azimuthal KM density [**FIG. 2**(b)] is also present because of the vortex phase. Therefore, the radial gradient of the axial KM density leads to the emergence of an azimuthal SAM density [**FIG. 2**(c)], whereas the radial gradient of the azimuthal KM density leads to the appearance of the axial SAM density [**FIG. 2**(d)], which indicates that unpolarized light also possesses a three-dimensional spin vector required to construct the variety of photonic spin topological defects. In addition, because the axial SAM density is related to the azimuthal KM density, it can be manipulated using several methods: 1. The magnitude and direction of the azimuthal KM density can be controlled by the vortex topological charge $\ell$, because the azimuthal KM density is proportional to $\ell$ [47, FIG. S5]; and 2. The magnitude of the axial SAM density is also related to the NA of objective lens because the orbit-to-spin AM conversion can be enhanced by enlarging the NA. For example, as the NA scales up from 0.2 to 0.8, the normalized axial SAM density also increases by two orders of magnitude [47, FIG. S4].

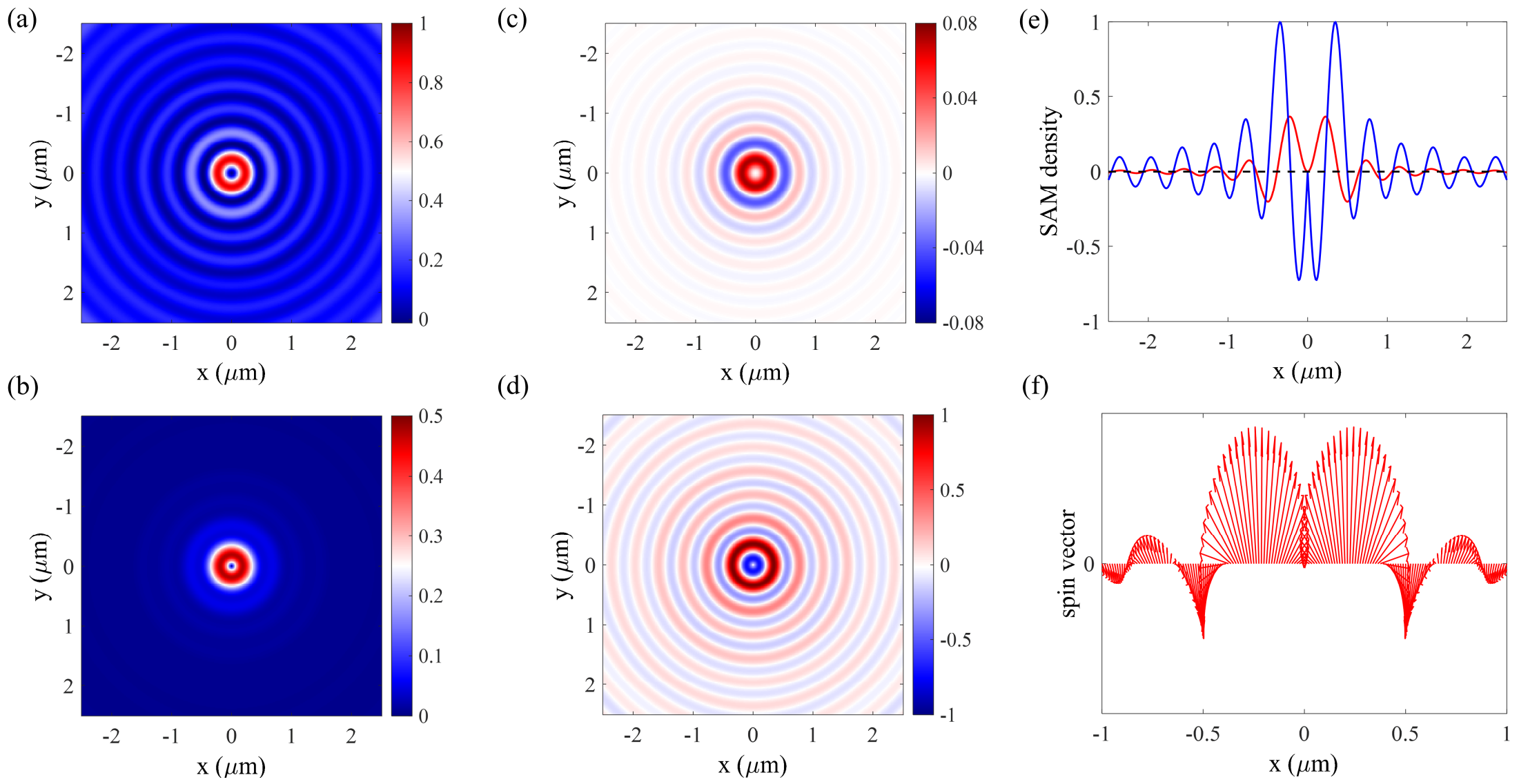

**FIG. 2.** Spin–momentum properties and spin textures of focused unpolarized vortex light using an objective lens of NA = 0.8. (a) Axial KM density and (b) azimuthal KM density of unpolarized light with vortex topological charge $\ell = 1$. According to the spin–momentum locking property, (c) the axial SAM density stems from the radial gradient of the azimuthal KM density, whereas (d) the azimuthal SAM density stems from the radial gradient of the axial KM density. From (e), the 1D contour of the SAM density (red line for the axial SAM density (the magnitude multiplied by 5) and blue line for the azimuthal SAM density), the null points of the axial SAM density and the azimuthal SAM density do not coincide, and thus the skyrmion number of (f) the spin texture formed by focused unpolarized vortex light is 0. The SAM density calculated from $\nabla\times\mathbf{P}/2k^2$ (blue line) and the coupling SAM density $\mathbf{S}_c$ (black line) indicate that, even for unpolarized light, the spin is locked with KM except for a tiny coupling term. Here, focused unpolarized vortex light is generated by passing linear polarized light through a RPP, which is subsequently focused by an objective lens.

On comparing topological spin textures constructed by the focused field of pure polarized vortex light [12–14], we find two primary differences between those textures formed by focused unpolarized vortex light and focused polarized light. First, the axial SAM density of focused unpolarized light exhibits a solid distribution when $\ell = \pm 2$, but has a donut shape for vortex topological charges taking other nonzero integer values [47, FIG. S5]. Second, the spin textures formed by focused unpolarized light is topological trivial, indicating that the skyrmion number is 0 [**FIG. 2**(e, f)]. This trivial spin texture originates from the spatial incoherence of focused unpolarized light, because the coherence vanishes among the focused components along the $\theta$-direction ($\theta$ the polar angle of spherical coordinates in focusing space). In addition, the coupling term $\mathbf{S}_c$ also affects the topological geometry of spin textures. For example, paraxial vortex light, which can be considered as weakly focused vortex light containing only one wavevector component, forms a half-skyrmion-like spin texture at the focal plane instead of a skyrmion-like spin texture [48].

To overcome the two limitations, we use a high-NA oil immersion objective lens to focus the unpolarized vortex light onto a metal film [**FIG. 3**(a)]. Only the TM polarized component is known to satisfy the wavevector matching condition and excite the surface plasmon (SP) mode at the metal/air interface [49]. In this instance, the spin–momentum relation for these SP vortex modes is expressible as [47, SM. IV]

$$\mathbf{S} = \frac{1}{2k^2}(\nabla \times \mathbf{P}) \tag{2}$$

because the metal film acts as a polarizer and the small coupling term vanishes. Equation (2) signifies that, even for the SP mode excited by unpolarized light, the total coherent SAM density is locked with the coherent KM density. Skyrmion-like spin textures can also form within focused unpolarized light. However, there are two differences between spin textures formed by polarized and unpolarized light. First, as with the SAM distributions of focused unpolarized light in free space, the axial SAM density displays a solid distribution when $\ell = \pm 2$ [**FIG. 3**(d, e) for $\ell = 2$], and forms a donut shape when the incident vortex takes topological charge of other nonzero integer values [**FIG. 3**(b, c) for $\ell = 1$ and **3**(f, g) for $\ell = 3$]. Second, from the 1D contours of SAM densities [47, FIG. S5], one finds that, for $\ell = 1$, the skyrmion-like spin texture forms in the center. The skyrmion number can be calculated from

$$\begin{aligned} n_{sk} &= \frac{1}{4\pi}\iint \mathbf{M}\cdot\left(\frac{\partial \mathbf{M}}{\partial x}\times\frac{\partial \mathbf{M}}{\partial y}\right)dxdy \\ &= -\frac{1}{2}\int_{r_n}^{r_{n+1}} M_r\left(M_r\frac{\partial M_z}{\partial r}-\frac{\partial M_r}{\partial r}M_z\right)dr \end{aligned}, \tag{3}$$

where $\mathbf{M} = [M_r, 0, M_z]^{\mathrm{T}}$ is the normalized spin vector ($\mathbf{M} = \mathbf{S}/|\mathbf{S}|$), and the superscript T signifies the transpose of matrix. For example, the skyrmion topological number of a spin texture from the center to the first approximate null point of the radial SAM density ($r_0 = 0$ and $r_1 = 495.2$nm) is $n_{sk} = 1$. Whereas for $\ell > 1$, no skyrmion-like spin texture is formed in the center. The reason lies in the expressions in [47, Eq. (S64-S66)], which reveal that the coefficients of the term containing $J_{\ell-1}$ and $J_{\ell+1}$ are non-vanishing and these two functions do not possess the same null points at small radial values. Here, the symbol J signifies a Bessel function of the first kind. Particularly, for large radial values, $J_{n-1}(\mathrm{r}) \approx \cos(r-(2n-1)\pi/4)$, and thus $J_{n+1}(\mathrm{r}) \approx \cos(r-(2n+3)\pi/4) = -\cos(r-(2n-1)\pi/4)$. In this instance, the null points of the terms containing $J_{\ell-1}$ and $J_{\ell+1}$ coincide and a skyrmion-like spin texture forms.

To validate the intriguing hidden spin–momentum locking and topological defects in a focused unpolarized field, we developed a scanning imaging system to map the out-of-plane SAM densities of the topological spin skyrmions. The experimental setup of the mapping the out-of-plane component of SAM densities of the focused unpolarized light with various vortex topological charges is given in [47, FIG. S7 and SM. V]. The incident light of 632.8nm-wavelength passes through a linear polarizer (LP), a half-wave plane (HWP) and a spatial light modulator (SLM) to generate a linear polarized light carrying the desired vortex phase. Then, the combination of two liquid crystal (LC) plates and a quarter-wave plate (QWP) is inserted to make sure that the polarization of light can be transformed into the arbitrary polarized state in the surface of Poincaré sphere. For example, the Jones matrix of linear polarized light is

$$\mathbf{E}_0 = \begin{bmatrix} 1 \\ 0 \end{bmatrix}.$$

The intersection angle between the fast axis of the first LC (LC1) and the polarization direction of the linear polarized light is 45°, and thus the Jones matrix of LC1 is

$$\mathbf{J}_{\mathrm{LC1}} = \cos\frac{\delta_1}{2}\begin{bmatrix} 1 & 0 \\ 0 & 1 \end{bmatrix} - i\sin\frac{\delta_1}{2}\begin{bmatrix} 0 & 1 \\ 1 & 0 \end{bmatrix}.$$

Here, $\delta_1$ represents the phase difference between the fast axis and the slow axis of LC1. i is the imaginary unit.

For the second LC (LC2), the intersection angle between the fast axis and the polarization direction of the linear polarized light is 0°, and thus its Jones matrix is

$$\mathbf{J}_{\mathrm{LC2}} = \cos\frac{\delta_2}{2}\begin{bmatrix}1 & 0\\ 0 & 1\end{bmatrix} - i\sin\frac{\delta_2}{2}\begin{bmatrix}1 & 0\\ 0 & -1\end{bmatrix}.$$

Here, $\delta_2$ represents the phase difference between the fast axis and the slow axis of LC2.

In addition, the intersection angle between the fast axis of QWP and the polarization direction of the linear polarized light is −45°, and thus its Jones matrix is

$$\mathbf{M} = \frac{1}{\sqrt{2}}\begin{bmatrix}1 & i\\ i & 1\end{bmatrix}.$$

From the expressions, we can calculate the Jones matrix of the linear polarized light passing through the combination of LC1, LC2 and QWP as

$$\begin{aligned}\mathbf{E} &= \mathbf{M}\mathbf{J}_{\mathrm{LC2}}\mathbf{J}_{\mathrm{LC1}}\mathbf{E}_0\\ &= \frac{1}{\sqrt{2}}\cos\left(\frac{\delta_1}{2}\right)e^{-i\frac{\delta_2}{2}}\begin{bmatrix}1\\ i\end{bmatrix} + \frac{1}{\sqrt{2}}\sin\left(\frac{\delta_1}{2}\right)e^{+i\frac{\delta_2}{2}}\begin{bmatrix}1\\ -i\end{bmatrix}.\end{aligned}$$

This expression is consistent with the polarization states in a Poincaré sphere [50]. To generate the polarization state covering any point on the surface of the Poincare sphere, we can control the phase delay of the LC device by the random voltage generated by the acquisition card. Finally, this Poincaré light beam is tightly focused by an oil immersion objective lens (Olympus, NA=1.49, 100×) to a PS nanoparticle with diameter 201nm sitting on a gold film (thickness: 45nm). The position of the PS particle can be precisely controlled by a piezo-stage (Physik Instrumente, P-545). The near-field signal that is scattered by the PS particle to an objective lens (Olympus, NA = 0.7, 60×) was split and then analyzed using a combination of a QWP and a LP to extract the individual circular polarization component ($I_{\mathrm{LCP}}$: left-handed circularly polarized component and $I_{\mathrm{RCP}}$: right-handed circularly polarized component) of the signal. Through the dipole theory [51], the horizontal components of electric field at each time can be reconstructed. These components were then directed to two photomultiplier tubes (PMTs) to measure the intensity information of the two signals. This then enables characterization of the out-of-plane SAM component (i.e., along the optical axis) of the focused beams by the relation [17]:

$$S_z = \frac{\varepsilon}{4\omega}\frac{k_{\perp,1}^2}{k_{z,1}^2}\left(I_{\mathrm{RCP}} - I_{\mathrm{LCP}}\right).$$

In this experiment, the experimental data of each pixel is the average of the collected results under the incident of multiple Poincaré polarized states. Here, we set the number of Poincaré polarized states as 50, and the pixel of image are 61*61 (the number of random voltages to be applied in the experiment is 50*61*61). The time interval between applying random voltages is 2ms, and we take about 12min to obtain an image of SAM density.

**FIG. 3**(b, d, f) plots the results calculated from theory of out-of-plane SAM densities for incident vortex light with vortex topological charge $\ell$ = 1, 2, 3. The corresponding experimental results are presented in **FIG. 3**(c, e, g). In addition, the experimental results of out-of-plane SAM densities for incident vortex light with vortex topological charge $\ell$ = 4 to 10 are given in [47, FIG. S8]. These results show that the distribution of the SAM density possesses a solid distribution at $\ell$ = 2 and in the other cases ($\ell$ a positive integer with $\ell \neq 2$) are hollow. The experimental results agree well with the theoretical results, thus demonstrating the validity of our theory.

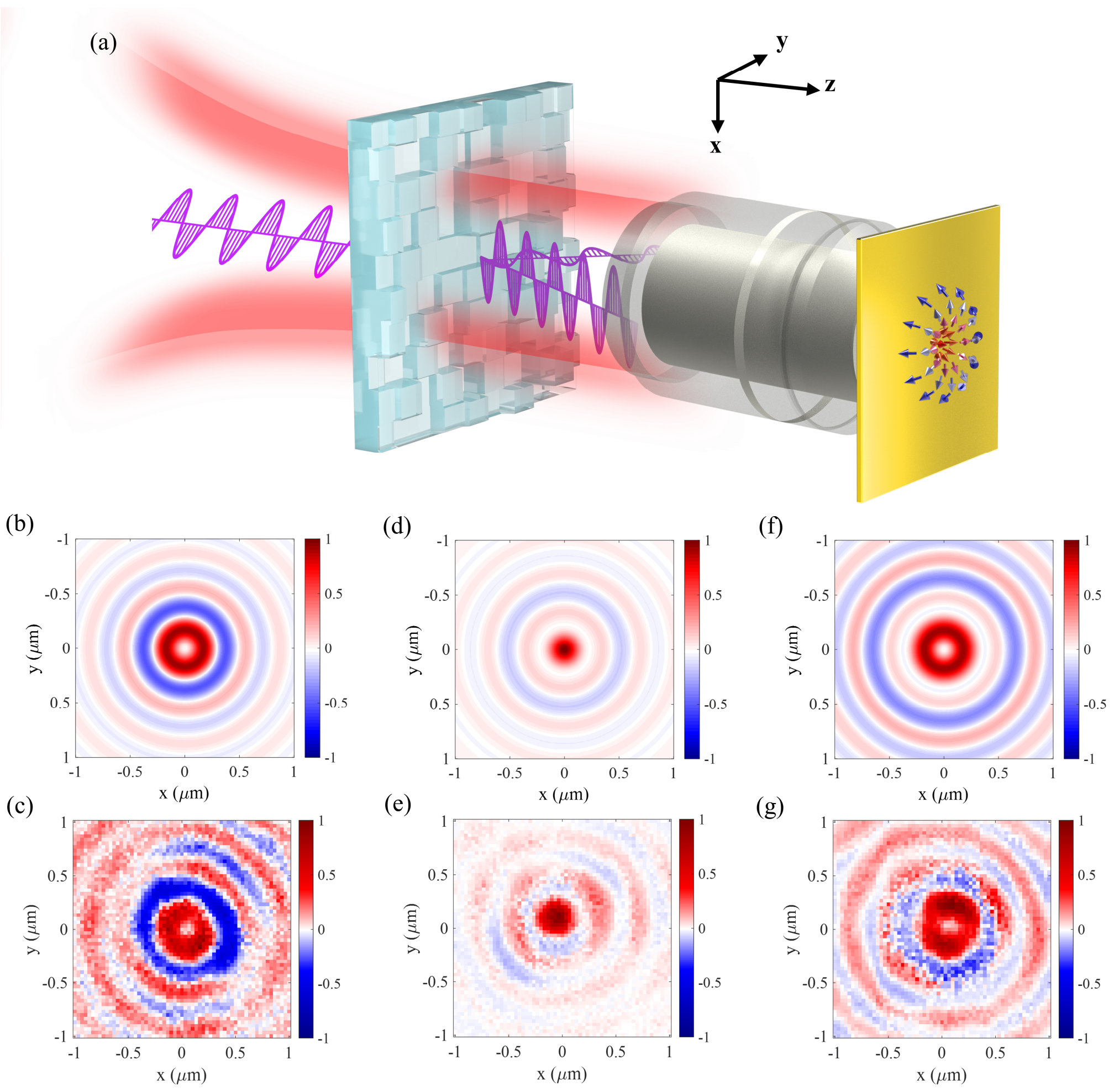

**FIG. 3.** Photonic spin skyrmions formed in focused unpolarized vortex light. (a) Schematic of the generation of photonic spin skyrmions in focused unpolarized vortex light. Linear polarized vortex light passes through a RPP and then is focused by an oil immersion objective lens with NA = 1.49 onto a metal surface. For an incident vortex of topological charge $\ell = 1$, (b) the theoretical and (c) the experimental axial SAM densities coincide strongly, verifying the photonic spin skyrmion formed by unpolarized light. (d, e) Same as (b, c) but with $\ell = 2$ and (f, g) Same as (b, c) but with $\ell = 3$. The wavelength is 632.8 nm; the thickness of the gold film is 45 nm.

## III. DISCUSSION

To summarize, we first discover and demonstrate theoretically and experimentally spin–momentum locking of unpolarized light, i.e., the coherent SAM density is locked with the coherent KM density except for a small coupling term. This term appears with the simultaneous presence of TM and TE polarized components in unpolarized light. Therefore, by employing a metal film acting as a polarizer, we conclude that the total SAM density is locked with the KM density. By adding a vortex phase to the incident light, skyrmion-like spin textures form at the metal/air interface. Interestingly, by comparing the skyrmion-like spin textures formed by focused polarized light, the distributions of spin textures formed by focused unpolarized light have two significant features: 1) the axial SAM density has a solid distribution when the vortex topological charge is $\ell = \pm 2$ and possesses a donut shape for other nonzero integers; and 2), apart from $\ell = \pm 1$, the skyrmion-like spin textures only form at large radial values whereas the skyrmion topological number is zero for spin textures in the center.

On the experimental side, we developed an in-house near-field optical scanning microscopic system to image the out-of-plane SAM densities of these topological spin textures. The experimental results match well with the theoretical results, thus validating our theory. The theory and technique proposed here can be applied to the fields of robust optical data encoding, encryption, and communications based on topological defects.

## ACKNOWLEDGMENTS

This work is supported by the Guangdong Major Project of Basic Research grant 2020B0301030009, the National Natural Science Foundation of China grants 12174266, 92250304, 61935013, 62075139, 61427819, 61622504, and 12004260, the Leadership of Guangdong province program grant 00201505.

## REFERENCES

*†These authors contributed equally to this work.*

*∗pittshiustc@gmail.com*

*∗lpdu@szu.edu.cn*

*∗xcyuan@szu.edu.cn*

[1] Bliokh, K. Y. Rodríguez-Fortuño, F. J., Nori, F., and Zayats, A. V. Spin–orbit interactions of light. Nat. Photon. **9**(12), 796–808 (2015).

[2] Aiello, A., Banzer, P., Neugebauer, M., and Leuchs, G. From transverse angular momentum to photonic wheels. Nat. Photon. **9**(12), 789–795 (2015).

[3] Shi, P., Yang, A., Meng, F., Chen, J., Zhang, Y., Xie, Z., Du, L., and Yuan, X. Optical near-field measurement for spin-orbit interaction of light. Prog. in Quantum Electron. **78**, 100341 (2021).

[4] Shi, P., Du, L., and Yuan, X. Spin photonics: from transverse spin to photonic skyrmions. Nanophotonics **10**, 3927-3943 (2021).

[5] Shen, Y. *et al.* Optical vortices 30 years on: OAM manipulation from topological charge to multiple singularities, Light: Science & Applications **8**, 90(2019).

[6] Bliokh, K. Y. and Bliokh, Y. P. Conservation of angular momentum, transverse shift, and spin Hall effect in reflection and refraction of an electromagnetic wave packet. Phys. Rev. Lett. **96**, 073903 (2006).

[7] Ling, X., Zhou, X., Huang, K., Liu, Y., Qiu, C.-W., Luo, H., and Wen, S. Recent advances in the spin Hall effect of light. Rep. Prog. Phys. **80**, 066401 (2017).

[8] Bliokh, K. Y., Niv, A., Kleiner, V., and Hasman, E. Geometrodynamics of spinning light. Nat. Photon. **2**, 748–753 (2008).

[9] Zhao, Y., Edgar, J. S., Jeffries, G. D. M., McGloin, D., and Chiu, D. T. Spin-to-orbital angular momentum conversion in a strongly focused optical beam. Phys. Rev. Lett. **99**, 073901 (2007).

[10] Neugebauer, M., Eismann, J. S., Bauer, T., and Banzer, P. Magnetic and electric transverse spin density of spatially confined light, Phys. Rev. X **8**, 021042 (2018).

[11] Bliokh, K. Y., Smirnova, D. & Nori, F. Quantum spin Hall effect of light. Science **348**, 1448–1451 (2015).

[12] Shi, P., Du, L., Li, C., Zayats, A. V., and Yuan, X. Transverse spin dynamics in structured electromagnetic guided waves. Proc. Natl. Acad. Sci. USA. **118**(6), e2018816118 (2021).

[13] Shi, P., Lei, X., Zhang, Q., Li, H., Du, L., and Yuan, X. Intrinsic spin-momentum dynamics of surface electromagnetic waves in dispersive interfaces, Phys. Rev. Lett. **128**, 213904 (2022).

[14] Shi, P., Yang, A., Yin, X., Du, L., Lei, X., and Yuan, X. Dynamical and topological properties of the spin angular momenta in general electromagnetic fields, Commun. Phys. **6**, 283 (2023).

[15] Meng, F., Yang, A., Du, K. *et al.* Measuring the magnetic topological spin structure of light using an anapole probe. Light Sci. Appl. **11**, 287 (2022).

[16] Dai, Y., Zhou, Z., Ghosh, A., Mong, R. S. K., Kubo, A., Huang, C.-B., and Petek, H. Plasmonic topological quasiparticle on the nanometre and femtosecond scales, Nature **588**, 616-619 (2020).

[17] Li, C., Shi, P., Du, L., and Yuan, X. Mapping the near-field spin angular momenta in the structured surface

plasmon polariton field, Nanoscale **12**, 13674-13679 (2020).

[18] Ghosh, A., Yang, S., Dai, Y., Zhou, Z., Wang, T., Huang, C., and Petek, H. A topological lattice of plasmonic merons, Appl. Phys. Rev. **8**, 041413 (2021).

[19] Shi, P., Du, L., and Yuan, X. Strong spin–orbit interaction of photonic skyrmions at the general optical interface, Nanophotonics **9**(15), 4619-4628 (2020).

[20] Dai, Y., Zhou, Z., Ghosh, A., Kapoor, K., Dąbrowski, M., Kubo, A., Huang, C.-B., and Petek, H. Ultrafast microscopy of a twisted plasmonic spin skyrmion, Appl. Phys. Rev. **9**, 011420 (2022).

[21] Lei, X., Yang, A., Shi, P., Xie, Z., Du, L., Zayats, A. V., and Yuan, X. Photonic spin lattices: symmetry constraints for skyrmion and meron topologies, Phys. Rev. Lett. **127**, 237403 (2021).

[22] Lu, C., Wang, B., Fang, X., Tsai, D. P., Zhu, W., Song, Q., Deng, X., He, T., Gong, X., Luo, H., Wang, Z., Dai, X., Shi, Y., and Cheng, X. Nanoparticle deep-subwavelength dynamics empowered by optical meron–antimeron topology, Nano Lett. (2023). DOI: 10.1021/acs.nanolett.3c03351

[23] Yu, X., Li, Y., Xu, B., Wang, X., Zhang, L., Chen, J., Lin, Z., and Chan, C.-T. Anomalous lateral optical force as a manifestation of the optical transverse spin, Laser Photonics Rev. **17**(9), 202300212 (2023).

[24] Shi, Y., Zhu, T., Liu, J., Tsai, D.-P., Zhang, H., Wang, S., Chan, C.-T., Wu, P.-C.., Zayats, A. V., Nori, F., and Liu, A.-Q. Stable optical lateral forces from inhomogeneities of the spin angular momentum, Sci. Adv. **8**(48), abn2291 (2022).

[25] Zhou, J. *et al.* Metasurface enabled quantum edge detection, Sci. Adv. **6**(51), eabc4385(2020).

[26] Zhou, J. *et al.* Optical edge detection based on high-efficiency dielectric metasurface, Proc. Natl. Acad. Sci. USA. **116**(23), 11137-11140(2019).

[27] Lei, X., Du, L., Yuan, X., and Zayats, A. V. Optical spin–orbit coupling in the presence of magnetization: photonic skyrmion interaction with magnetic domains, Nanophotonics **10**(14), 3667–3675(2021).

[28] Yang, A. et al. Spin-manipulated photonic skyrmion-pair for pico-metric displacement sensing, Adv. Sci. **10**(12), 2205249 (2023).

[29] Gu, M., Zhang, Q., and Lamon, S. Nanomaterials for optical data storage, Nat. Rev. Mater. **1**, 16070 (2016).

[30] Yang, H., He, P., Ou, K. et al. Angular momentum holography via a minimalist metasurface for optical nested encryption. Light Sci. Appl. **12**, 79 (2023).

[31] Liu, S., Wang, X., Ni, J., Cao, Y., Li, J., Wang, C., Hu, Y., Chu, J., and Wu, D. Optical encryption in the photonic orbital angular momentum dimension via direct-laser-writing 3D chiral metahelices. Nano Lett. **23**(6) 2304–2311 (2023).

[32] Fang, X., Ren, H. & Gu, M. Orbital angular momentum holography for high-security encryption. Nat. Photon. **14**, 102–108 (2020).

[33] Ren, H., Fang, X., Jang, J. et al. Complex-amplitude metasurface-based orbital angular momentum holography in momentum space. Nat. Nanotechnol. **15**, 948–955 (2020).

[34] Qu, G., Yang, W., Song, Q. et al. Reprogrammable meta-hologram for optical encryption. Nat. Commun. **11**, 5484 (2020).

[35] Alfalou, A. and Brosseau, C. Optical image compression and encryption methods. Adv. Opt. Photon. **1**, 589-636 (2009).

[36] Ouyang, X., Xu, Y., Xian, M. et al. Synthetic helical dichroism for six-dimensional optical orbital angular momentum multiplexing. Nat. Photon. **15**, 901–907 (2021).

[37] Lei, T., Zhang, M., Li, Y. et al. Massive individual orbital angular momentum channels for multiplexing enabled by Dammann gratings. Light Sci. Appl. **4**, e257 (2015).

[38] Wang, J., Yang, J. Y., Fazal, I. et al. Terabit free-space data transmission employing orbital angular momentum multiplexing. Nat. Photon. **6**, 488–496 (2012).

[39] Eismann, J.S., Nicholls, L.H., Roth, D.J. et al. Transverse spinning of unpolarized light. Nat. Photon. **15**, 156–161 (2021).

[40] Forbes, Kayn A. Optical helicity of unpolarized light,

Phys. Rev. A 105, 023524 (2022).

[41] Chen, Y., Wang, F., Dong, Z., Cai, Y., Norrman, A., Gil, J. J., Friberg, A. T., and Setälä, T. Structure of transverse spin in focused random light, Phys. Rev. A **104**, 013516 (2021).

[42] Chen, Y., Norrman, A., Ponomarenko, S. A., and Friberg, A. T. Spin density in partially coherent surface-plasmon-polariton vortex fields, Phys. Rev. A **103**, 063511 (2021).

[43] He, C., Shen, Y. & Forbes, A. Towards higher-dimensional structured light. Light Sci. Appl. **11**, 205 (2022).

[44] Liu, C., Zhang, S., Maier, S. A., and Ren, H. Disorder-induced topological state transition in the optical skyrmion family. Phys. Rev. Lett. **129**, 267401 (2022).

[45] Shi, P., Du, L., Li, M., and Yuan, X. Symmetry-protected photonic chiral spin textures by spin-orbit coupling, Laser Photonics Rev. **15**, 202000554 (2021).

[46] Shi, P., Zhang, Q., and Yuan, X. Topological state transitions in electromagnetic topological defects. https://doi.org/10.48550/arXiv.2306.12024 (2023).

[47] See Supplemental Material for discussions about I. field, polarization, and coherent properties of the unpolarized light, II. momentum and spin densities of the focused unpolarized light, III. intrinsic spin-momentum locking for the focused unpolarized light, IV. topological defects in the focused unpolarized light, and V. experimental setup and measurement of SAM density of the focused unpolarized vortex light at the metal/air interface.

[48] Shi, P., Li, H., Du, L., and Yuan, X. Spin-momentum properties in the paraxial optical systems. ACS Photonics **10**, 2332 (2023).

[49] Shen, Y., Zhang, Q., Shi, P., Du, L., Yuan, X., and Zayats, A. V. Topological quasiparticles of light: Optical skyrmions and beyond. Nat. Photon. 18, 15 (2024).

[50] Milione, G., Sztul, H. I., Nolan, D. A., and Alfano, R. R. Higher-Order Poincaré Sphere, Stokes Parameters, and the Angular Momentum of Light. Phys. Rev. Lett. **107**, 053601 (2011).

[51] Novotny, L., Hecht, B., and Keller, O. *Principles of nano-optics* (Cambridge University Press, 2012).

[52] Bliokh, K. Y., Bekshaev, A. Y., and Nori, F. Optical momentum, spin, and angular momentum in dispersive media, Phys. Rev. Lett. **119**, 073901(2017).

[53] Zayats, A. V., Smolyaninov, I. I., and Maradudin, A. A. Nano-optics of surface plasmon polaritons, Phys. Rep. **408**, 131 (2005).

Supplementary Materials for

# "The hidden spin-momentum locking and topological defects in unpolarized light fields"

Peng Shi[1,#,*], Min Li[1,#], Xinxin Gou[1,#], Luping Du[1,*], Aiping Yang[2], and Xiaocong Yuan[1,3,*]

[1]*Nanophotonics Research Centre, Institute of Microscale Optoelectronics & State Key Laboratory of Radio Frequency Heterogeneous Integration, Shenzhen University, Shenzhen 518060, China*

[2]*Research Institute of Interdisciplinary Science & School of Materials Science and Engineering, Dongguan University of Technology, Dongguan 523808, China*

[3]*Zhejiang Lab, Research Center for Humanoid Sensing, Research Institute of Intelligent Sensing, Hangzhou 311100, China*

[#]These authors contribute equally to the work.

[*]Authors to whom correspondence should be addressed: *pittshiustc@gmail.com*, *lpdu@szu.edu.cn* and *xcyuan@szu.edu.cn*.

## I. Field, polarization, and coherent properties of the unpolarized light

In the manuscript, we only consider a time-harmonic monochromatic electromagnetic (EM) field (**E**: electric field; **H**: magnetic field) with angular frequency $\omega$. The SI units are used through the main text and supplementary information. We assume that the unpolarized light is constructed by a linear polarized light passing through a random phase plate (RPP) whose amplitude transmissions and phase delays for the $x$- and $y$- polarized components can be tuned separately. The schematic diagram of a linear polarized light under such a random perturbation can be found in FIG. S1.

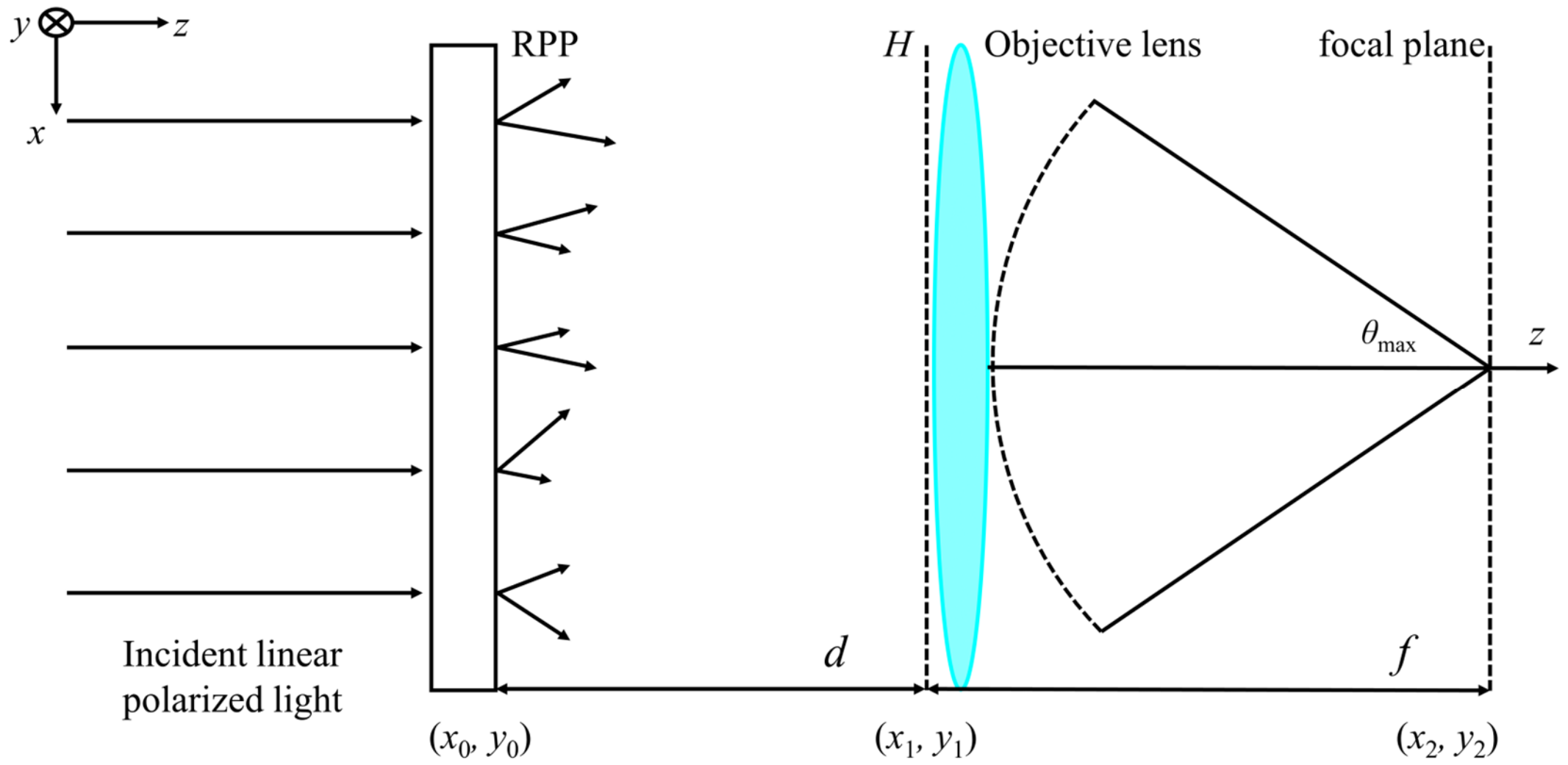

**FIG. 1.** Schematic diagram of generation of a unpolarized light with a linear polarized incident light and a RPP. The unpolarized light is then focused by an objective lens with focal length $f$ and numerical aperture NA = sin$\theta_{max}$.

In Cartesian coordinates ($x$, $y$, $z$), the transmission coefficients of the RPP can be expressed as

$$\mathbf{T}_0\left(x_0, y_0, t\right)=\sum_{n=1}^{N}\sum_{m=1}^{N}\delta\left(x_0-u_n\right)\delta\left(y_0-v_m\right)\begin{bmatrix} a_{nm}\cos\theta_{nm}\hat{\mathbf{x}}_0 \\ \chi_{nm}a_{nm}\sin\theta_{nm}\hat{\mathbf{y}}_0 \end{bmatrix}e^{i\varphi_{nm}} . \quad \text{(S1)}$$

Here, $\varphi_{nm}$ is the instantaneous phase of the RPP at the position ($u_n$, $v_m$) and the time $t$. The number of sample point is $N$ in each the $x$- and $y$-direction. $a_{nm}\cos\theta_{nm}$ and $a_{nm}\sin\theta_{nm}$ are the instantaneous transmission coefficients of the RPP for the $x$- and $y$-polarized components, respectively. $\chi_{nm} = \exp(i\phi_{nm})$ with $\chi_{nm}^*\chi_{nm} = 1$ is the instantaneous phase difference between the $x$- and $y$-polarized components. All the parameters $a_{nm}$, $\theta_{nm}$, $\chi_{nm}$, and $\phi_{nm}$ depend on the position ($u_n$, $v_m$) and the time $t$. The transverse field components in the plane ($x_1$, $y_1$, $z$) can be expressed as

$$\begin{aligned}\mathbf{E}_{\perp 1}\left(x_1, y_1, t\right)&=\iint_{-\infty}^{\infty}\mathbf{T}_0\left(x_0, y_0, t\right)\exp\left\{i\frac{\pi}{\lambda d}\left[\left(x_1-x_0\right)^2+\left(y_1-y_0\right)^2\right]\right\}dx_0dy_0 \\ &=\sum_{n=1}^{N}\sum_{m=1}^{N}a_{nm}\begin{bmatrix}\cos\theta_{nm}\hat{\mathbf{x}}_1 \\ \chi_{nm}\sin\theta_{nm}\hat{\mathbf{y}}_1\end{bmatrix}\exp\left[i\varphi_{nm}\right]\exp\left\{i\frac{\pi}{\lambda d}\left[\left(x_1-u_n\right)^2+\left(y_1-v_m\right)^2\right]\right\}\end{aligned} . \quad \text{(S2)}$$

The Jones matrix of the linear polarized light passing through a RPP can be expressed as

$$\mathbf{J}=\begin{bmatrix}E_{x1}\\E_{y1}\end{bmatrix}=\begin{bmatrix}\sum_{n=1}^{N}\sum_{m=1}^{N}a_{nm}\cos\theta_{nm}\exp[i\varphi_{nm}]\exp\left\{i\frac{\pi}{\lambda d}\left[(\rho_1\cos\phi_1-u_n)^2+(\rho_1\sin\phi_1-v_m)^2\right]\right\}\\\sum_{n=1}^{N}\sum_{m=1}^{N}a_{nm}\chi_{nm}\sin\theta_{nm}\exp[i\varphi_{nm}]\exp\left\{i\frac{\pi}{\lambda d}\left[(\rho_1\cos\phi_1-u_n)^2+(\rho_1\sin\phi_1-v_m)^2\right]\right\}\end{bmatrix}. \quad \text{(S3)}$$

Here, there are $\rho_1=\sqrt{x_1^2+y_1^2}=fsin\theta_1$ with $f$ the focal length of objective lens and $\phi_1=\arctan y_1/x_1$. It is worth noting that the time integral of $E_{x1}$ or $E_{y1}$ is zero because the phase function $\varphi_{nm}$, the transmission coefficients $a_{nm}$ and $\theta_{nm}$, and phase difference $\chi_{nm}$ are completely random and incoherent. Therefore, the degree of polarization of the unpolarized light is zero.

This can be also verified by calculating the coherence matrix. For example, the coherence matrix element $\overline{E_x^*E_x}$ of the linear polarized light passing through a RPP can be calculated as

$$\begin{aligned}\overline{E_x^*E_x}&=\sum_{n=1}^{N}\sum_{m=1}^{N}\sum_{i=1}^{N}\sum_{j=1}^{N}a_{nm}^*a_{ij}\cos\theta_{nm}\cos\theta_{ij}e^{-i(\varphi_{nm}-\varphi_{ij})}e^{-i\frac{\pi}{\lambda d}\left[(x_1-u_n)^2+(y_1-v_m)^2-(x_1-u_i)^2-(y_1-v_j)^2\right]}\\&=\sum_{n=1}^{N}\sum_{m=1}^{N}a_{nm}^*a_{nm}\cos\theta_{nm}\cos\theta_{nm}+\sum_{n\neq i}^{N}\sum_{m\neq j}^{N}a_{nm}^*a_{ij}\cos\theta_{nm}\cos\theta_{ij}e^{-i(\varphi_{nm}-\varphi_{ij})}e^{-i\frac{\pi}{\lambda d}\left[(x_1-u_n)^2+(y_1-v_m)^2-(x_1-u_i)^2-(y_1-v_j)^2\right]}\end{aligned}. \quad \text{(S4)}$$

Considering the self-correlation properties of this linear polarized light passing through a RPP, we can obtain several expressions

$$\sum_{n=1}^{N}\sum_{m=1}^{N}\cos\theta_{nm}\cos\theta_{nm}=\frac{N^2}{2}, \quad \text{(S5)}$$

$$\sum_{n=1}^{N}\sum_{m=1}^{N}\sin\theta_{nm}\sin\theta_{nm}=\frac{N^2}{2}, \quad \text{(S6)}$$

$$\sum_{n=1}^{N}\sum_{m=1}^{N}\sin\theta_{nm}\cos\theta_{nm}=0, \quad \text{(S7)}$$

$$\sum_{n\neq i}^{N}\sum_{m\neq j}^{N}\cos\theta_{nm}\cos\theta_{nm}=0, \quad \text{(S8)}$$

$$\sum_{n\neq i}^{N}\sum_{m\neq j}^{N}\sin\theta_{nm}\sin\theta_{nm}=0, \quad \text{(S9)}$$

$$\sum_{n\neq i}^{N}\sum_{m\neq j}^{N}\sin\theta_{nm}\cos\theta_{nm}=0, \quad \text{(S10)}$$

and

$$N^2a_{nm}^*a_{nm}=1. \quad \text{(S11)}$$

Therefore, the coherence matrix elements are calculated as

$$\overline{E_x^*E_x}=\frac{1}{2},\ \overline{E_y^*E_y}=\frac{1}{2},\ \overline{E_x^*E_y}=0,\text{ and }\overline{E_y^*E_x}=0. \quad \text{(S12)}$$

By employing the expression (S12), we obtain that the coherence matrix is

$$\overline{\mathbb{R}}=\overline{\begin{bmatrix}E_x\\E_y\end{bmatrix}\begin{bmatrix}E_x^* & E_y^*\end{bmatrix}}=\begin{bmatrix}\overline{E_xE_x^*} & \overline{E_xE_y^*}\\\overline{E_yE_x^*} & \overline{E_yE_y^*}\end{bmatrix}=\frac{1}{2}\begin{bmatrix}1 & 0\\0 & 1\end{bmatrix}, \quad \text{(S13)}$$

and the degree of polarization of this linear polarized light passing through a RPP is

$$P=\left[1-\frac{4\det\overline{\mathbb{R}}}{\left(Tr\overline{\mathbb{R}}\right)^2}\right]^{1/2}=\left[1-\frac{4\left\{\left(\overline{E_xE_x^*}\right)\left(\overline{E_yE_y^*}\right)-\left(\overline{E_xE_y^*}\right)\left(\overline{E_yE_x^*}\right)\right\}}{\left(\overline{E_xE_x^*}+\overline{E_yE_y^*}\right)^2}\right]^{1/2}=0\,, \tag{S14}$$

and the complex degree of coherence is

$$\mu_{12}=\frac{\overline{E_yE_x^*}}{\sqrt{\left(\overline{E_xE_x^*}\right)\left(\overline{E_yE_y^*}\right)}}=0\,. \tag{S15}$$

The Equation (S14) also represents the degree of polarization of the linear polarized light passing through a RPP is zero, which is consistent with the vanishing of the time integral of the Jones matrix elements.

## II. Momentum and spin densities of the focused unpolarized light

To consider the focusing properties of this unpolarized light, we first transform the electric field from Cartesian coordinates ($x_1$, $y_1$, $z$) to cylindrical coordinates ($\rho_1$, $\phi_1$, $z$) by the coordinate transformations $x_1=\rho_1\cos\phi_1$ and $y_1=\rho_1\sin\phi_1$:

$$\begin{bmatrix}E_{\rho 1}\\ E_{\phi 1}\end{bmatrix}=\begin{bmatrix}\sum_{n=1}^{N}\sum_{m=1}^{N}a_{nm}\left[+\cos\theta_{nm}\cos\phi_1+\chi_{nm}\sin\theta_{nm}\sin\phi_1\right]e^{i\varphi_{nm}}e^{i\frac{\pi}{\lambda d}\left[\left(\rho_1\cos\phi_1-u_n\right)^2+\left(\rho_1\sin\phi_1-v_m\right)^2\right]}\hat{\boldsymbol{\rho}}_1\\ \sum_{n=1}^{N}\sum_{m=1}^{N}a_{nm}\left[-\cos\theta_{nm}\sin\phi_1+\chi_{nm}\sin\theta_{nm}\cos\phi_1\right]e^{i\varphi_{nm}}e^{i\frac{\pi}{\lambda d}\left[\left(\rho_1\cos\phi_1-u_n\right)^2+\left(\rho_1\sin\phi_1-v_m\right)^2\right]}\hat{\boldsymbol{\phi}}_1\end{bmatrix}. \tag{S16}$$

Then, the focused electric and magnetic fields can be calculated as

$$\mathbf{E}_2\left(r_2,\varphi_2,z_2\right)=\int_0^{\theta_{\max}}\int_0^{2\pi}\sqrt{\cos\theta_1}\boldsymbol{\Gamma}_1^e\left(\theta_1,\phi_1\right)\cdot e^{ik\left[z_2\cos\theta_1+r_2\sin\theta_1\cos\left(\phi_1-\varphi_2\right)\right]}\sin\theta_1 d\phi_1 d\theta_1\,, \tag{S17}$$

and

$$\mathbf{H}_2\left(r_2,\varphi_2,z_2\right)=\eta A\int_0^{\theta_{\max}}\int_0^{2\pi}\sqrt{\cos\theta_1}\boldsymbol{\Gamma}_1^h\left(\theta_1,\phi_1\right)\cdot e^{ik\left[z_2\cos\theta_1+r_2\sin\theta_1\cos\left(\phi_1-\varphi_2\right)\right]}\sin\theta_1 d\phi_1 d\theta_1\,, \tag{S18}$$

respectively. Here, the constant $A=\frac{ikfe^{-ikf}}{2\pi}\sqrt{\frac{n_i\cos\theta_1}{n_o}}$, $\eta=k/\omega\mu$ and the integral quantities are

$$\boldsymbol{\Gamma}_1^e\left(\theta_1,\phi_1\right)=\begin{pmatrix}\Gamma_{x1}^e\hat{\mathbf{x}}_1\\ \Gamma_{y1}^e\hat{\mathbf{y}}_1\\ \Gamma_{z1}^e\hat{\mathbf{z}}_1\end{pmatrix}=\begin{pmatrix}\left[+t_p\left(\theta_1\right)\cos\theta_1\cos\phi_1E_{\rho 1}-t_s\left(\theta_1\right)\sin\phi_1E_{\phi 1}\right]\hat{\mathbf{x}}_1\\ \left[+t_p\left(\theta_1\right)\cos\theta_1\sin\phi_1E_{\rho 1}+t_s\left(\theta_1\right)\cos\phi_1E_{\phi 1}\right]\hat{\mathbf{y}}_1\\ \left[-t_p\left(\theta_1\right)\sin\theta_1E_{\rho 1}\right]\hat{\mathbf{z}}_1\end{pmatrix} \tag{S19}$$

and

$$\boldsymbol{\Gamma}_1^h\left(\theta_1,\phi_1\right)=\begin{pmatrix}\Gamma_{x1}^h\hat{\mathbf{x}}_1\\ \Gamma_{y1}^h\hat{\mathbf{y}}_1\\ \Gamma_{z1}^h\hat{\mathbf{z}}_1\end{pmatrix}=\begin{pmatrix}\left[-t_p\left(\theta_1\right)\sin\phi_1E_{\rho 1}-t_s\left(\theta_1\right)\cos\theta_1\cos\phi_1E_{\phi 1}\right]\hat{\mathbf{x}}_1\\ \left[+t_p\left(\theta_1\right)\cos\phi_1E_{\rho 1}-t_s\left(\theta_1\right)\cos\theta_1\sin\phi_1E_{\phi 1}\right]\hat{\mathbf{y}}_1\\ \left[+t_s\left(\theta_1\right)\sin\theta_1E_{\phi 1}\right]\hat{\mathbf{z}}_1\end{pmatrix} \tag{S20}$$

for Equation (S17) and Equation (S18), respectively. $t_p(\theta_1)$ and $t_s(\theta_1)$ denote the Fresnel transmission coefficients of $p$-polarization (transvers magnetic) and $s$-polarization (transvers electric) components, respectively, and we ignore the ($\theta_1$)-dependent in the following expressions. In the focusing space, the field components are labeled by the subscript 2 in Cartesian coordinates ($x_2$, $y_2$, $z$). The field components can be transformed from Cartesian coordinates to cylindrical coordinates ($r_2$, $\varphi_2$, $z$) by the transformations $x_2 = r_2\cos\varphi_2$ and $y_2 = r_2\sin\varphi_2$. To further simplify the expressions, we employ the Bessel's integration to the focused electric/magnetic field and obtain

$$\mathbf{E}_2\left(r_2,\varphi_2,z_2\right) = A\int_0^{\theta_{\max}} \sqrt{\cos\theta_1}\boldsymbol{\Sigma}^e \cdot e^{ikz_2\cos\theta_1}\sin\theta_1 d\theta_1 , \tag{S21}$$

and

$$\mathbf{H}_2\left(r_2,\varphi_2,z_2\right) = \eta A\int_0^{\theta_{\max}} \sqrt{\cos\theta_1}\boldsymbol{\Sigma}^h \cdot e^{ikz_2\cos\theta_1}\sin\theta_1 d\theta_1 , \tag{S22}$$

in which the integral quantities are

$$\boldsymbol{\Sigma}^e = \begin{pmatrix} \Sigma_x^e\hat{\mathbf{x}}_1 \\ \Sigma_y^e\hat{\mathbf{y}}_1 \\ \Sigma_z^e\hat{\mathbf{z}}_1 \end{pmatrix} \tag{S23}$$

and

$$\boldsymbol{\Sigma}^h = \begin{pmatrix} \Sigma_x^h\hat{\mathbf{x}}_1 \\ \Sigma_y^h\hat{\mathbf{y}}_1 \\ \Sigma_z^h\hat{\mathbf{z}}_1 \end{pmatrix}. \tag{S24}$$

Here, there are

$$\Sigma_x^e = \pi\sum_{n=1}^{N}\sum_{m=1}^{N} a_{nm}\begin{bmatrix} +t_p\cos\theta_1\begin{bmatrix} +\cos\theta_{nm}J_0\left(\Pi_{nm}\right) \\ -\left(\cos\theta_{nm}\cos 2\Theta_{nm} + \chi_{nm}\sin\theta_{nm}\sin 2\Theta_{nm}\right)J_2\left(\Pi_{nm}\right) \end{bmatrix} \\ -t_s\begin{bmatrix} -\cos\theta_{nm}J_0\left(\Pi_{nm}\right) \\ -\left(\cos\theta_{nm}\cos 2\Theta_{nm} + \chi_{nm}\sin\theta_{nm}\sin 2\Theta_{nm}\right)J_2\left(\Pi_{nm}\right) \end{bmatrix} \end{bmatrix} e^{i\left[\varphi_{nm}+\frac{\pi}{\lambda d}\left(\rho_1^2+R_{nm}^2\right)\right]} , \tag{S25}$$

$$\Sigma_y^e = \pi\sum_{n=1}^{N}\sum_{m=1}^{N} a_{nm}\begin{bmatrix} +t_p\cos\theta_1\begin{bmatrix} +\chi_{nm}\sin\theta_{nm}J_0\left(\Pi_{nm}\right) \\ -\left(\cos\theta_{nm}\sin 2\Theta_{nm} - \chi_{nm}\sin\theta_{nm}\cos 2\Theta_{nm}\right)J_2\left(\Pi_{nm}\right) \end{bmatrix} \\ +t_s\begin{bmatrix} +\chi_{nm}\sin\theta_{nm}J_0\left(\Pi_{nm}\right) \\ +\left(\cos\theta_{nm}\sin 2\Theta_{nm} - \chi_{nm}\sin\theta_{nm}\cos 2\Theta_{nm}\right)J_2\left(\Pi_{nm}\right) \end{bmatrix} \end{bmatrix} e^{i\left[\varphi_{nm}+\frac{\pi}{\lambda d}\left(\rho_1^2+R_{nm}^2\right)\right]} , \tag{S26}$$

$$\Sigma_z^e = \pi\sum_{n=1}^{N}\sum_{m=1}^{N} a_{nm}\left[-2it_p\sin\theta_1\left[+\cos\theta_{nm}\cos\Theta_{nm} + \chi_{nm}\sin\theta_{nm}\sin\Theta_{nm}\right]J_1\left(\Pi_{nm}\right)\right]e^{i\left[\varphi_{nm}+\frac{\pi}{\lambda d}\left(\rho_1^2+R_{nm}^2\right)\right]} , \tag{S27}$$

$$\Sigma_x^h = \pi\sum_{n=1}^{N}\sum_{m=1}^{N} a_{nm}\begin{bmatrix} -t_p\begin{bmatrix} +\chi_{nm}\sin\theta_{nm}J_0\left(\Pi_{nm}\right) \\ -\left(\cos\theta_{nm}\sin 2\Theta_{nm} - \chi_{nm}\sin\theta_{nm}\cos 2\Theta_{nm}\right)J_2\left(\Pi_{nm}\right) \end{bmatrix} \\ -t_s\cos\theta_1\begin{bmatrix} +\chi_{nm}\sin\theta_{nm}J_0\left(\Pi_{nm}\right) \\ +\left(\cos\theta_{nm}\sin 2\Theta_{nm} - \chi_{nm}\sin\theta_{nm}\cos 2\Theta_{nm}\right)J_2\left(\Pi_{nm}\right) \end{bmatrix} \end{bmatrix} e^{i\left[\varphi_{nm}+\frac{\pi}{\lambda d}\left(\rho_1^2+R_{nm}^2\right)\right]} , \tag{S28}$$

$$\Sigma_y^h = \pi\sum_{n=1}^{N}\sum_{m=1}^{N} a_{nm}\begin{bmatrix} +t_p\begin{bmatrix} +\cos\theta_{nm}J_0\left(\Pi_{nm}\right) \\ -\left(\cos\theta_{nm}\cos 2\Theta_{nm} + \chi_{nm}\sin\theta_{nm}\sin 2\Theta_{nm}\right)J_2\left(\Pi_{nm}\right) \end{bmatrix} \\ -t_s\cos\theta_1\begin{bmatrix} -\cos\theta_{nm}J_0\left(\Pi_{nm}\right) \\ -\left(\cos\theta_{nm}\cos 2\Theta_{nm} + \chi_{nm}\sin\theta_{nm}\sin 2\Theta_{nm}\right)J_2\left(\Pi_{nm}\right) \end{bmatrix} \end{bmatrix} e^{i\left[\varphi_{nm}+\frac{\pi}{\lambda d}\left(\rho_1^2+R_{nm}^2\right)\right]} , \tag{S29}$$

and

$$\Sigma_z^h = \pi \sum_{n=1}^{N}\sum_{m=1}^{N} a_{nm}\left[-2it_s \sin\theta_1\left[+\cos\theta_{nm}\sin\Theta_{nm} - \chi_{nm}\sin\theta_{nm}\cos\Theta_{nm}\right]J_1\left(\Pi_{nm}\right)\right]e^{i\left[\varphi_{nm}+\frac{\pi}{\lambda d}\left(\rho_1^2+R_{nm}^2\right)\right]}. \quad \text{(S30)}$$

We use the substitutions

$$u_n = R_{nm}\cos\zeta_{nm}, \quad \text{(S31)}$$

$$v_m = R_{nm}\sin\zeta_{nm}, \quad \text{(S32)}$$

$$\Pi_{nm} = \sqrt{\left(-\frac{2\pi\rho_1 R_{nm}}{\lambda d}\cos\zeta_{nm} + kr_2\sin\theta_1\cos\varphi_2\right)^2 + \left(-\frac{2\pi\rho_1 R_{nm}}{\lambda d}\sin\zeta_{nm} + kr_2\sin\theta_1\sin\varphi_2\right)^2}, \quad \text{(S33)}$$

and

$$\Theta_{nm} = \arctan\frac{\left(-\frac{2\pi\rho_1 R_{nm}}{\lambda d}\sin\zeta_{nm} + kr_2\sin\theta_1\sin\varphi_2\right)}{\left(-\frac{2\pi\rho_1 R_{nm}}{\lambda d}\cos\zeta_{nm} + kr_2\sin\theta_1\cos\varphi_2\right)}. \quad \text{(S34)}$$

It is worth noting that the spatial coherence of RPP vanishes, and thus the quantities $R_{nm}$ and $\zeta_{nm}$ can be considered as the stochastic disturbance on the radius and phase of focused field and the stochastic disturbance can be cancelled when we analyse the coherent momentum and angular momentum of the focused field of the unpolarized light.

For a time-harmonic monochromatic EM field in a lossless isotropic medium with electric permittivity $\varepsilon$ and magnetic permeability $\mu$, the kinetic-Abraham-Poynting momentum (KM) density **P** and the spin angular momentum (SAM) density **S** are [52]:

$$\mathbf{P} = \frac{1}{2c^2}\mathrm{Re}\left\{\mathbf{E}_2^* \times \mathbf{H}_2\right\}, \quad \text{(S35)}$$

and

$$\mathbf{S} = \frac{1}{4\omega}\mathrm{Im}\left\{\varepsilon\left(\mathbf{E}_2^* \times \mathbf{E}_2\right) + \mu\left(\mathbf{H}_2^* \times \mathbf{H}_2\right)\right\}. \quad \text{(S36)}$$

Here, $c$ is the speed of light in vacuum and we do not consider the dispersive effect, which will induce an additional group term in the permittivity and permeability [52]. By using the techniques as Equations (S4-S11), we can obtain the expressions of the KM density and SAM density the focused field of the unpolarized light as

$$P_r = \frac{\pi^2\eta|A|^2}{2c^2}\int_0^{\theta_{\max}}\int_0^{\theta_{\max}} \mathrm{Im}\left\{\begin{array}{l}\sqrt{\cos\theta_{1,1}\cos\theta_{1,2}} \\ \times\left[\begin{array}{l}+\sin\theta_{1,1}J_0\left(\Pi_2\right)J_1\left(\Pi_1\right)t_{p,1}^*\left(t_{p,2}+t_{s,2}\cos\theta_{1,2}\right) \\ -\sin\theta_{1,2}J_0\left(\Pi_1\right)J_1\left(\Pi_2\right)t_{s,2}\left(t_{s,1}^*+t_{p,1}^*\cos\theta_{1,1}\right) \\ -\sin\theta_{1,1}J_1\left(\Pi_1\right)J_2\left(\Pi_2\right)t_{p,1}^*\left(t_{p,2}-t_{s,2}\cos\theta_{1,2}\right) \\ +\sin\theta_{1,2}J_1\left(\Pi_2\right)J_2\left(\Pi_1\right)t_{s,2}\left(t_{s,1}^*-t_{p,1}^*\cos\theta_{1,1}\right)\end{array}\right] \\ \times\exp\left\{-i\frac{\pi}{\lambda d}\left[\rho_{1,1}^2-\rho_{1,2}^2\right]\right\}e^{-ikz\left(\cos\theta_{1,1}-\cos\theta_{1,2}\right)}\sin\theta_{1,1}\sin\theta_{1,2}\end{array}\right\}d\theta_{1,1}d\theta_{1,2}, \quad \text{(S37)}$$

$$P_\varphi = 0\,, \tag{S38}$$

$$P_z = \frac{\pi^2\eta|A|^2}{2c^2}\int_0^{\theta_{\max}}\int_0^{\theta_{\max}} \mathrm{Re}\left\{\begin{array}{l}\sqrt{\cos\theta_{1,1}\cos\theta_{1,2}} \\ \times\left[\begin{array}{l}+J_0(\Pi_1)J_0(\Pi_2)\left(t_{s,1}^*+t_{p,1}^*\cos\theta_{1,1}\right)\left(t_{p,2}+t_{s,2}\cos\theta_{1,2}\right) \\ -J_2(\Pi_1)J_2(\Pi_2)\left(t_{s,1}^*-t_{p,1}^*\cos\theta_{1,1}\right)\left(t_{p,2}-t_{s,2}\cos\theta_{1,2}\right)\end{array}\right] \\ \times\exp\left\{-i\frac{\pi}{\lambda d}\left[\rho_{1,1}^2-\rho_{1,2}^2\right]\right\}e^{-ikz\left(\cos\theta_{1,1}-\cos\theta_{1,2}\right)}\sin\theta_{1,1}\sin\theta_{1,2}\end{array}\right\}d\theta_{1,1}d\theta_{1,2}\,, \tag{S39}$$

and

$$S_r = 0\,, \tag{S40}$$

$$S_\varphi = \frac{\varepsilon\pi^2|A|^2}{4\omega}\int_0^{\theta_{\max}}\int_0^{\theta_{\max}} \mathrm{Re}\left\{\begin{array}{l}\sqrt{\cos\theta_{1,1}\cos\theta_{1,2}} \\ \times\left[\begin{array}{l}+\sin\theta_{1,1}J_0(\Pi_2)J_1(\Pi_1)t_{p,1}^*\left(t_{s,2}+t_{p,2}\cos\theta_{1,2}\right) \\ +\sin\theta_{1,1}J_0(\Pi_2)J_1(\Pi_1)t_{s,1}^*\left(t_{p,2}+t_{s,2}\cos\theta_{1,2}\right) \\ +\sin\theta_{1,2}J_0(\Pi_1)J_1(\Pi_2)t_{p,2}\left(t_{s,1}^*+t_{p,1}^*\cos\theta_{1,1}\right) \\ +\sin\theta_{1,2}J_0(\Pi_1)J_1(\Pi_2)t_{s,2}\left(t_{p,1}^*+t_{s,1}^*\cos\theta_{1,1}\right) \\ +\sin\theta_{1,1}J_1(\Pi_1)J_2(\Pi_2)t_{p,1}^*\left(t_{s,2}-t_{p,2}\cos\theta_{1,2}\right) \\ +\sin\theta_{1,1}J_1(\Pi_1)J_2(\Pi_2)t_{s,1}^*\left(t_{p,2}-t_{s,2}\cos\theta_{1,2}\right) \\ +\sin\theta_{1,2}J_1(\Pi_2)J_2(\Pi_1)t_{p,2}\left(t_{s,1}^*-t_{p,1}^*\cos\theta_{1,1}\right) \\ +\sin\theta_{1,2}J_1(\Pi_2)J_2(\Pi_1)t_{s,2}\left(t_{p,1}^*-t_{s,1}^*\cos\theta_{1,1}\right)\end{array}\right] \\ \times\exp\left\{-i\frac{\pi}{\lambda d}\left[\rho_{1,1}^2-\rho_{1,2}^2\right]\right\}e^{-ikz\left(\cos\theta_{1,1}-\cos\theta_{1,2}\right)}\sin\theta_{1,1}\sin\theta_{1,2}\end{array}\right\}d\theta_{1,1}d\theta_{1,2}\,, \tag{S41}$$

$$S_z = 0\,. \tag{S42}$$

Here, by considering the spatial incoherence, there are

$$\Pi_l = kr_2\sin\theta_{1,l}\,, \tag{S43}$$

and

$$\Theta_l = \varphi_2\,. \tag{S44}$$

The subscript $l$ = 1, 2 after the comma represents the different integral quantity. It can be observed that the KM density contains the axial component originated from the propagation property of the unpolarized light and the radial component originated from the focusing or divergence property in the front or afterwards the focal plane. Particularly, the focused unpolarized light field possesses the SAM density, and the direction of this SAM density is transverse to that of KM density, which is similar to the transverse spin of the Gaussian focusing field.

## III. Intrinsic spin-momentum locking for the focused unpolarized light

Spin-momentum locking (SML) is a fundamental dynamical property of transverse spin for a polarized EM field [12-14]. This property can also be extended to several classical fields, such as, acoustic wave and gravity water wave, etc. Here, we demonstrate that, even for a unpolarized light, this SML relationship also satisfies. Here, for simplicity, we consider the focusing of unpolarized light in the free space with the relative electric permittivity $\varepsilon_r = 1$ and the relative magnetic permeability $\mu_r = 1$. Therefore, the terms $n_i/n_o = 1$, $t_s = 1$ and $t_p = 1$. The vorticity of the KM density is

$$
\begin{aligned}
(\nabla\times\mathbf{P})_r &= \frac{1}{r_2}\frac{\partial P_z}{\partial\varphi_2} - \frac{\partial P_\varphi}{\partial z_2} \\
&= \frac{\pi^2\eta|A|^2}{2c^2 r_2}\int_0^{\theta_{\max}}\int_0^{\theta_{\max}} \frac{\partial\,\mathrm{Re}}{\partial\varphi_2}\left\{\begin{array}{l}
\sqrt{\cos\theta_{1,1}\cos\theta_{1,2}} \\
\times\left[\begin{array}{l}
+J_0(\Pi_1)J_0(\Pi_2)\left(t_{s,1}^*+t_{p,1}^*\cos\theta_{1,1}\right)\left(t_{p,2}+t_{s,2}\cos\theta_{1,2}\right) \\
-J_2(\Pi_1)J_2(\Pi_2)\left(t_{s,1}^*-t_{p,1}^*\cos\theta_{1,1}\right)\left(t_{p,2}-t_{s,2}\cos\theta_{1,2}\right)
\end{array}\right] \\
\times\exp\left\{-i\frac{\pi}{\lambda d}\left[\rho_{1,1}^2-\rho_{1,2}^2\right]\right\}e^{-ikz\left(\cos\theta_{1,1}-\cos\theta_{1,2}\right)}\sin{}^{1}\theta_1\sin\theta_{1,2}
\end{array}\right\}d\theta_{1,1}d\theta_{1,2} = 0
\end{aligned}, \quad \text{(S45)}
$$

$$
\begin{aligned}
(\nabla\times\mathbf{P})_\varphi &= \frac{\partial P_r}{\partial z_2} - \frac{\partial P_z}{\partial r_2} \\
&= \frac{\pi^2 k\eta|A|^2}{2c^2}\int_0^{\theta_{\max}}\int_0^{\theta_{\max}} \mathrm{Re}\left\{\begin{array}{l}
\sqrt{\cos\theta_{1,1}\cos\theta_{1,2}} \\
\times\left[\begin{array}{l}
-\sin\theta_{1,1}J_0(\Pi_2)J_1(\Pi_1)t_{p,1}^*\left(t_{p,2}+t_{s,2}\cos\theta_{1,2}\right)\left(\cos\theta_{1,1}-\cos\theta_{1,2}\right) \\
+\sin\theta_{1,2}J_0(\Pi_1)J_1(\Pi_2)t_{s,2}\left(t_{s,1}^*+t_{p,1}^*\cos\theta_{1,1}\right)\left(\cos\theta_{1,1}-\cos\theta_{1,2}\right) \\
+\sin\theta_{1,1}J_1(\Pi_1)J_2(\Pi_2)t_{p,1}^*\left(t_{p,2}-t_{s,2}\cos\theta_{1,2}\right)\left(\cos\theta_{1,1}-\cos\theta_{1,2}\right) \\
-\sin\theta_{1,2}J_1(\Pi_2)J_2(\Pi_1)t_{s,2}\left(t_{s,1}^*-t_{p,1}^*\cos\theta_{1,1}\right)\left(\cos\theta_{1,1}-\cos\theta_{1,2}\right) \\
+\sin\theta_{1,1}J_0(\Pi_2)J_1(\Pi_1)\left(t_{s,1}^*+t_{p,1}^*\cos\theta_{1,1}\right)\left(t_{p,2}+t_{s,2}\cos\theta_{1,2}\right) \\
+\sin\theta_{1,2}J_0(\Pi_1)J_1(\Pi_2)\left(t_{s,1}^*+t_{p,1}^*\cos\theta_{1,1}\right)\left(t_{p,2}+t_{s,2}\cos\theta_{1,2}\right) \\
+\sin\theta_{1,1}J_2'(\Pi_1)J_2(\Pi_2)\left(t_{s,1}^*-t_{p,1}^*\cos\theta_{1,1}\right)\left(t_{p,2}-t_{s,2}\cos\theta_{1,2}\right) \\
+\sin\theta_{1,2}J_2(\Pi_1)J_2'(\Pi_2)\left(t_{s,1}^*-t_{p,1}^*\cos\theta_{1,1}\right)\left(t_{p,2}-t_{s,2}\cos\theta_{1,2}\right)
\end{array}\right] \\
\times\exp\left\{-i\frac{\pi}{\lambda d}\left[\rho_{1,1}^2-\rho_{1,2}^2\right]\right\}e^{-ikz\left(\cos\theta_{1,1}-\cos\theta_{1,2}\right)}\sin\theta_{1,1}\sin\theta_{1,2}
\end{array}\right\}d\theta_{1,1}d\theta_{1,2}
\end{aligned}, \quad \text{(S46)}
$$

$$
\begin{aligned}
(\nabla\times\mathbf{P})_z &= \frac{1}{r_2}\frac{\partial r_2 P_\varphi}{\partial r_2} - \frac{1}{r_2}\frac{\partial P_r}{\partial\varphi_2} \\
&= -\frac{\pi^2\eta|A|^2}{2c^2 r_2}\int_0^{\theta_{\max}}\int_0^{\theta_{\max}} \frac{\partial\,\mathrm{Re}}{\partial\varphi_2}\left\{\begin{array}{l}
\sqrt{\cos\theta_{1,1}\cos\theta_{1,2}} \\
\times\left[\begin{array}{l}
+\sin\theta_{1,1}J_0(\Pi_2)J_1(\Pi_1)t_{p,1}^*\left(t_{p,2}+t_{s,2}\cos\theta_{1,2}\right) \\
-\sin\theta_{1,2}J_0(\Pi_1)J_1(\Pi_2)t_{s,2}\left(t_{s,1}^*+t_{p,1}^*\cos\theta_{1,1}\right) \\
-\sin\theta_{1,1}J_1(\Pi_1)J_2(\Pi_2)t_{p,1}^*\left(t_{p,2}-t_{s,2}\cos\theta_{1,2}\right) \\
+\sin\theta_{1,2}J_1(\Pi_2)J_2(\Pi_1)t_{s,2}\left(t_{s,1}^*-t_{p,1}^*\cos\theta_{1,1}\right)
\end{array}\right] \\
\times\exp\left\{-i\frac{\pi}{\lambda d}\left[\rho_{1,1}^2-\rho_{1,2}^2\right]\right\}e^{-ikz\left(\cos\theta_{1,1}-\cos\theta_{1,2}\right)}\sin{}^{1}\theta_1\sin\theta_{1,2}
\end{array}\right\}d\theta_{1,1}d\theta_{1,2} = 0
\end{aligned}, \quad \text{(S47)}
$$

It can be observed that only the azimuthal component of vorticity of KM density does not vanish. The azimuthal component of vorticity of KM density can be transformed into

$$(\nabla\times\mathbf{P})_\varphi=\frac{\partial P_r}{\partial z_2}-\frac{\partial P_z}{\partial r_2}$$

$$=\frac{\pi^2 k\eta|A|^2}{2c^2}\int_0^{\theta_{\max}}\int_0^{\theta_{\max}}\mathrm{Re}\left\{\begin{array}{l}\sqrt{\cos\theta_{1,1}\cos\theta_{1,2}}\\ \left[\begin{array}{l}+\sin\theta_{1,1}J_0(\Pi_2)J_1(\Pi_1)\left(t_{s,1}^*+t_{p,1}^*\cos\theta_{1,2}\right)\left(t_{p,2}+t_{s,2}\cos\theta_{1,2}\right)\\ +\sin^2\theta_1 J_0(\Pi_1)J_1(\Pi_2)\left(t_{s,1}^*+t_{p,1}^*\cos\theta_{1,1}\right)\left(t_{p,2}+t_{s,2}\cos\theta_{1,1}\right)\\ +\sin\theta_{1,1}J_1(\Pi_1)J_2(\Pi_2)\left(t_{s,1}^*-t_{p,1}^*\cos\theta_{1,2}\right)\left(t_{p,2}-t_{s,2}\cos\theta_{1,2}\right)\\ +\sin^2\theta_1 J_2(\Pi_1)J_1(\Pi_2)\left(t_{s,1}^*-t_{p,1}^*\cos\theta_{1,1}\right)\left(t_{p,2}-t_{s,2}\cos\theta_{1,1}\right)\\ -\frac{4}{kr_2}J_2(\Pi_1)J_2(\Pi_2)\left(t_{s,1}^*-t_{p,1}^*\cos\theta_{1,1}\right)\left(t_{p,2}-t_{s,2}\cos\theta_{1,2}\right)\end{array}\right]\\ \times\exp\left\{-i\frac{\pi}{\lambda d}\left[\rho_{1,1}^2-\rho_{1,2}^2\right]\right\}e^{-ikz\left(\cos\theta_{1,1}-\cos\theta_{1,2}\right)}\sin\theta_{1,1}\sin\theta_{1,2}\end{array}\right\}d\theta_{1,1}d\theta_{1,2}\ . \quad \text{(S48)}$$

Furthermore, if we employ the approximation of Bessel function at large value, the last term in expression (S44) can be ignored. Then, comparing the vorticity of KM density with the SAM density, we obtain that

$$S_\varphi-\frac{1}{2k^2}(\nabla\times\mathbf{P})_\varphi$$

$$=\frac{\varepsilon\pi^2|A|^2}{4\omega}\int_0^{\theta_{\max}}\int_0^{\theta_{\max}}\mathrm{Re}\left\{\begin{array}{l}t_{p,1}^*t_{s,2}\sqrt{\cos\theta_{1,1}\cos\theta_{1,2}}\left[\begin{array}{l}+\sin\theta_{1,1}\sin^2\theta_{1,2}J_0(\Pi_2)J_1(\Pi_1)\\ +\sin\theta_{1,2}\sin^2\theta_{1,1}J_0(\Pi_1)J_1(\Pi_2)\\ +\sin\theta_{1,1}\sin^2\theta_{1,2}J_1(\Pi_1)J_2(\Pi_2)\\ +\sin\theta_{1,2}\sin^2\theta_{1,1}J_1(\Pi_2)J_2(\Pi_1)\end{array}\right]\\ \times\exp\left\{-i\frac{\pi}{\lambda d}\left[\rho_{1,1}^2-\rho_{1,2}^2\right]\right\}e^{-ikz\left(\cos\theta_{1,1}-\cos\theta_{1,2}\right)}\sin\theta_{1,1}\sin\theta_{1,2}\end{array}\right\}d\theta_{1,1}d\theta_{1,2}\ . \quad \text{(S49)}$$

Therefore, the spin-momentum locking of the focused unpolarized light can be re-expressed as

$$\mathbf{S}=\frac{1}{2k^2}(\nabla\times\mathbf{P})+\mathbf{S}_c\,. \quad \text{(S50)}$$

The term $\mathbf{S}_c$ is named from the coupling between the *s*-polarization and *p*-polarization components, because all the terms in $\mathbf{S}_c$ depend on the $t_p^* t_s$ representing the coupling between Fresnel transmission coefficients for the *s*-polarization and *p*-polarization components. From the Equation (S50), we can realize that: 1. The coupling term $\mathbf{S}_c$ is primarily related to $\sin\theta_{1,1}$ and $\sin\theta_{1,2}$, and thus the coupling term can be ignored when the unpolarized light is weakly focused (In the instance, $\sin\theta\approx 0$, as shown in **FIG. S2**); 2. For the focused unpolarized light, the SAM density (as shown in **FIG. S2**(b-c), **S2**(f-g), **S2**(j-k)) is also locked with the KM density (as shown in **FIG. S2**(a), **S2**(e), **S2**(i)) except for a small coupling term $\mathbf{S}_c$ (as shown in **FIG. S2**(d), **S2**(h), **S2**(l)). Therefore, when the propagating direction given by KM density is inverted, the SAM density is also reversed. This property is completely consistent with that of a general polarized EM field [14], and moreover, the coupling term is helicity-independent. From the Equation (S49), we can conclude that the spin properties of focused unpolarized light can be tuned by the energy flow (proportional to KM density) and focusing property (weakly or tightly focusing by NA). Furthermore, we can give the quantitative expression for the KM density and SAM density of focusing unpolarized light, and thus the Belinfante spin momentum ($\mathbf{P}_s$ = $\nabla\times\mathbf{S}/2$), and imaginary momentum $\mathbf{P}_{Im} = \mathrm{Im}(\mathbf{E}_2^* \times \mathbf{H}_2)$ can be also obtained, which can be utilized to perform the calculation of the focusing unpolarized light based lateral optical force for chiral sorting applications.

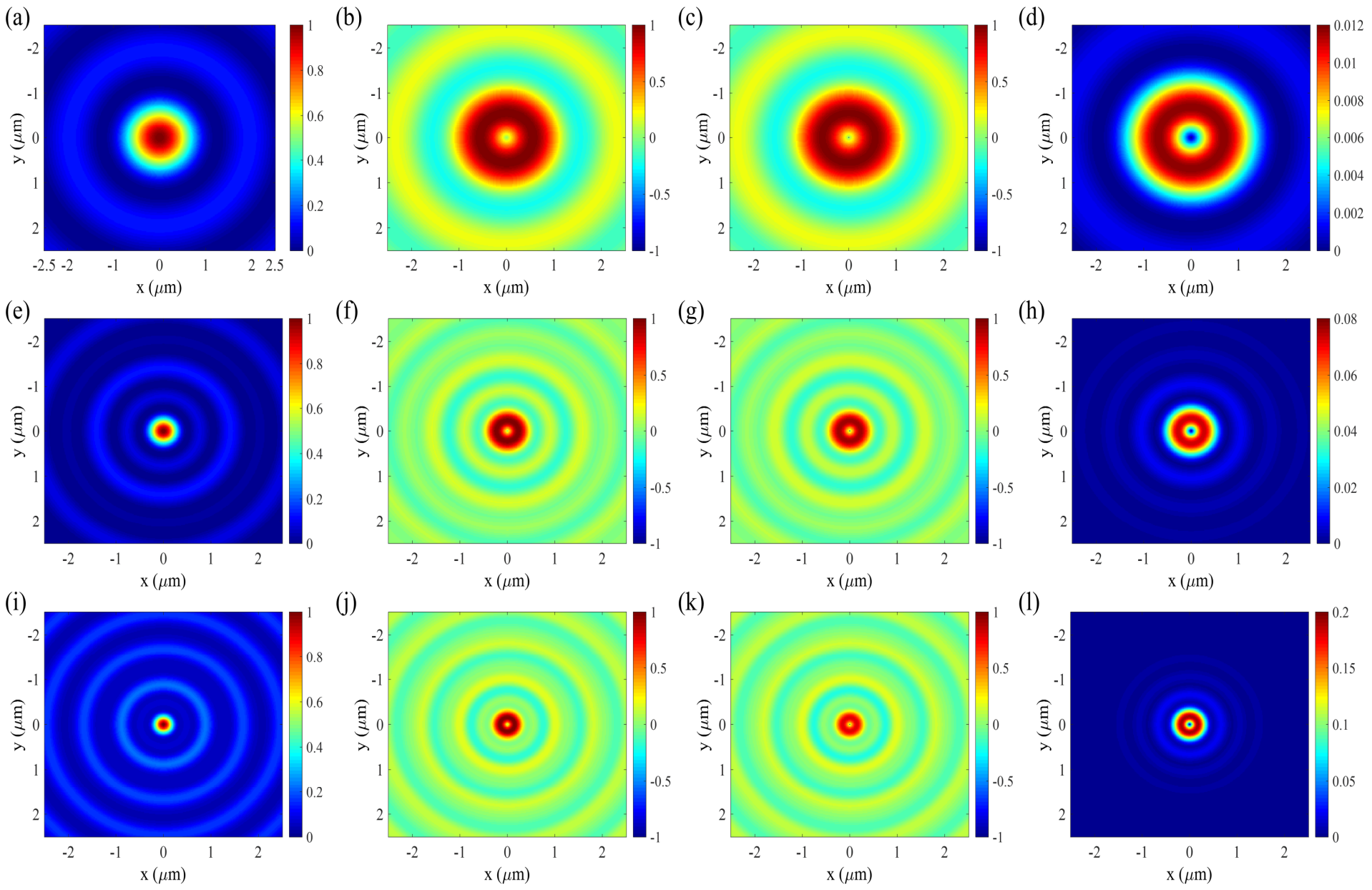

**FIG. 2.** The KM and SAM properties of the focused unpolarized light at the focal plane. When the NA of objective lens is 0.2, (a) The axial KM density, (b) the azimuthal SAM density, (c) the vorticity of KM density, and (d) the coupling SAM density $\mathbf{S}_c$ given by the difference between the total SAM density and the vorticity of KM density $\nabla\times\mathbf{P}/2k^2$ are shown. In this instance, because the angle of convergence is small, the coupling SAM density $\mathbf{S}_c$ is also much small according to Equation (S49). (e-h) Same with (a-d) but with NA = 0.5. (i-l) Same with (a-d) but with NA = 0.8. It can be observed that, as the NA = sin$\theta_{\max}$ enlarges, the coupling SAM density $\mathbf{S}_c$ increases according to Equation (S49). Nevertheless, the coupling SAM density $\mathbf{S}_c$ is still much smaller than the vorticity of KM density $\nabla\times\mathbf{P}/2k^2$. Therefore, we can conclude that the SAM density is locking with the KM density satisfying the right-hand rule.

## IV. Topological defects in the focused unpolarized light

In the former section, we demonstrate the SML property of focused unpolarized light constructed by a linear polarized plane wave passing through a RPP. For this focused unpolarized light, only azimuthal SAM density exists, and no topological spin texture presents. In this section, we consider a class of focused unpolarized light induced by a linear polarized vortex light passing through a combination of a RPP, as shown in **FIG. S3**.

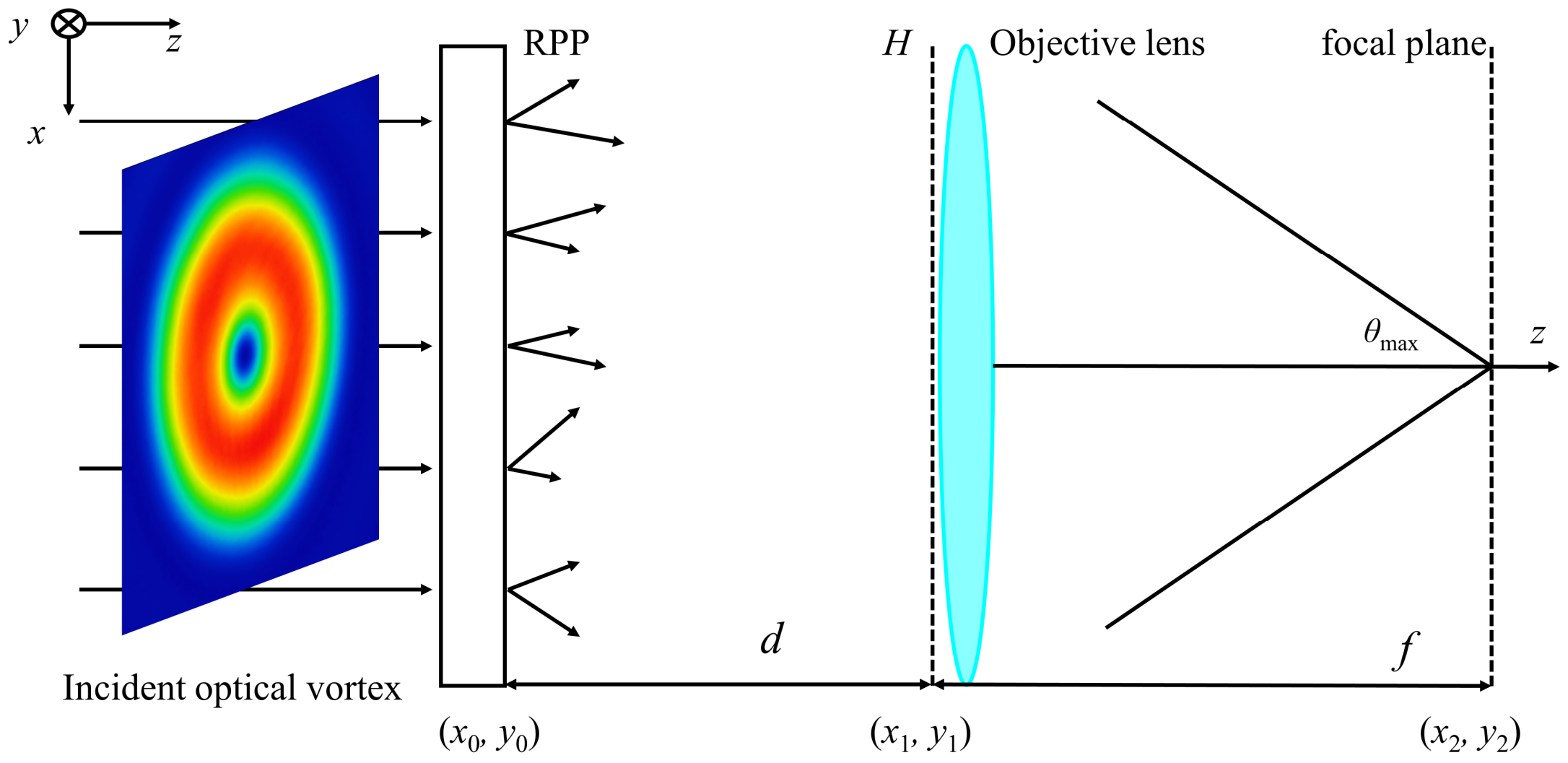

**FIG. 3.** Schematic diagram of the generation of unpolarized light by a linear polarized vortex light passing through a RPP.

The transverse field components in the plane ($x_1$, $y_1$, $z$) can be expressed by

$$\begin{aligned}\mathbf{E}_{\perp 1}(x_1,y_1,t) &= e^{i\ell\phi_1}\iint_{-\infty}^{\infty}\mathbf{T}_0(x_0,y_0,t)\exp\left\{i\frac{\pi}{\lambda d}\left[(x_1-x_0)^2+(y_1-y_0)^2\right]\right\}dx_0dy_0 \\ &= e^{i\ell\phi_1}\sum_{n=1}^{N}\sum_{m=1}^{N}a_{nm}\begin{bmatrix}\cos\theta_{nm}\hat{\mathbf{x}}_1\\ \chi_{nm}\sin\theta_{nm}\hat{\mathbf{y}}_1\end{bmatrix}\exp\left[i\varphi_{nm}\right]\exp\left\{i\frac{\pi}{\lambda d}\left[(x_1-u_n)^2+(y_1-v_m)^2\right]\right\}\end{aligned}. \tag{S51}$$

Although there is an additional vortex phase with topological charge $\ell$ appearance in the transverse electric field components, this vortex phase is hardly to be recognized because it is covered up by the random perturbation. The coherence matrix of this unpolarized light constructed by a linear polarized light passing through a combination of a RPP and a vortex phase plate are

$$\overline{E_x^*E_x}=\frac{1}{2},\quad \overline{E_y^*E_y}=\frac{1}{2},\quad \overline{E_x^*E_y}=0,\ \text{and}\ \overline{E_y^*E_x}=0. \tag{S52}$$

The coherence matrix and the degree of polarization for this unpolarized vortex light are

$$\overline{\mathbb{R}}=\overline{\begin{bmatrix}E_x\\E_y\end{bmatrix}\begin{bmatrix}E_x^* & E_y^*\end{bmatrix}}=\begin{bmatrix}\overline{E_xE_x^*} & \overline{E_xE_y^*}\\ \overline{E_yE_x^*} & \overline{E_yE_y^*}\end{bmatrix}=\frac{1}{2}\begin{bmatrix}1&0\\0&1\end{bmatrix}, \tag{S53}$$

and

$$P=\left[1-\frac{4\det\overline{\mathbb{R}}}{\left(Tr\overline{\mathbb{R}}\right)^2}\right]^{1/2}=\left[1-\frac{4\left\{\left(\overline{E_xE_x^*}\right)\left(\overline{E_yE_y^*}\right)-\left(\overline{E_xE_y^*}\right)\left(\overline{E_yE_x^*}\right)\right\}}{\left(\overline{E_xE_x^*}+\overline{E_yE_y^*}\right)^2}\right]^{1/2}=0, \tag{S54}$$

respectively. Therefore, one cannot recognize the signal of vortex phase merely from the field information of this unpolarized light field.

In the following, we calculate the KM density and the SAM density of the focused field of this unpolarized light. The KM density components in cylindrical coordinates ($r_1$, $\phi_1$, z) are

$$P_r = \frac{\pi^2 \eta |A|^2}{4c^2} \int_0^{\theta_{\max}} \int_0^{\theta_{\max}} \mathrm{Re} \left\{ \begin{array}{l} \sqrt{\cos^* \theta_{1,1} \cos\theta_{1,2}} \sin\theta_{1,1} \sin\theta_{1,2} \\ \times \left[ \begin{array}{l} -i \sin\theta_{1,1} \begin{bmatrix} +J_\ell(\Pi_2) J_{\ell+1}(\Pi_1) \\ -J_\ell(\Pi_2) J_{\ell-1}(\Pi_1) \end{bmatrix} t_{p,1}^* \left( t_{p,2} + t_{s,2} \cos\theta_{1,2} \right) \\ -i \sin\theta_{1,1} \begin{bmatrix} -J_{\ell+1}(\Pi_1) J_{\ell+2}(\Pi_2) \\ +J_{\ell-1}(\Pi_1) J_{\ell-2}(\Pi_2) \end{bmatrix} t_{p,1}^* \left( t_{p,2} - t_{s,2} \cos\theta_{1,2} \right) \\ +i \sin\theta_{1,2} \begin{bmatrix} +J_\ell(\Pi_1) J_{\ell+1}(\Pi_2) \\ -J_\ell(\Pi_1) J_{\ell-1}(\Pi_2) \end{bmatrix} t_{s,2} \left( t_{s,1}^* + t_{p,1}^* \cos^* \theta_{1,1} \right) \\ +i \sin\theta_{1,2} \begin{bmatrix} -J_{\ell+1}(\Pi_2) J_{\ell+2}(\Pi_1) \\ +J_{\ell-1}(\Pi_2) J_{\ell-2}(\Pi_1) \end{bmatrix} t_{s,2} \left( t_{s,1}^* - t_{p,1}^* \cos^* \theta_{1,1} \right) \end{array} \right] \\ \times \exp\left\{ -i \frac{\pi}{\lambda d} \left[ \rho_{1,2}^2 - \rho_{1,2}^2 \right] \right\} e^{-ikz\left( \cos^* \theta_{1,1} - \cos\theta_{1,2} \right)} \end{array} \right\} d\theta_{1,1} d\theta_{1,2} \text{ , (S55)}$$

$$P_\varphi = \frac{\pi^2 \eta |A|^2}{4c^2} \int_0^{\theta_{\max}} \int_0^{\theta_{\max}} \mathrm{Re} \left\{ \begin{array}{l} \sqrt{\cos^* \theta_{1,1} \cos\theta_{1,2}} \sin\theta_{1,1} \sin\theta_{1,2} \\ \times \left[ \begin{array}{l} +\sin\theta_{1,1} \begin{bmatrix} +J_\ell(\Pi_2) J_{\ell+1}(\Pi_1) \\ +J_\ell(\Pi_2) J_{\ell-1}(\Pi_1) \end{bmatrix} t_{p,1}^* \left( t_{p,2} + t_{s,2} \cos\theta_{1,2} \right) \\ +\sin\theta_{1,1} \begin{bmatrix} +J_{\ell+1}(\Pi_1) J_{\ell+2}(\Pi_2) \\ +J_{\ell-1}(\Pi_1) J_{\ell-2}(\Pi_2) \end{bmatrix} t_{p,1}^* \left( t_{p,2} - t_{s,2} \cos\theta_{1,2} \right) \\ +\sin\theta_{1,2} \begin{bmatrix} +J_\ell(\Pi_1) J_{\ell+1}(\Pi_2) \\ +J_\ell(\Pi_1) J_{\ell-1}(\Pi_2) \end{bmatrix} t_{s,2} \left( t_{s,1}^* + t_{p,1}^* \cos^* \theta_{1,1} \right) \\ +\sin\theta_{1,2} \begin{bmatrix} +J_{\ell+1}(\Pi_2) J_{\ell+2}(\Pi_1) \\ +J_{\ell-1}(\Pi_2) J_{\ell-2}(\Pi_1) \end{bmatrix} t_{s,2} \left( t_{s,1}^* - t_{p,1}^* \cos^* \theta_{1,1} \right) \end{array} \right] \\ \times \exp\left\{ -i \frac{\pi}{\lambda d} \left[ \rho_{1,1}^2 - \rho_{1,2}^2 \right] \right\} e^{-ikz\left( \cos^* \theta_{1,1} - \cos\theta_{1,2} \right)} \end{array} \right\} d\theta_{1,1} d\theta_{1,2} \text{ , (S56)}$$

$$P_z = \frac{\pi^2 \eta |A|^2}{4c^2} \int_0^{\theta_{\max}} \int_0^{\theta_{\max}} \mathrm{Re} \left\{ \begin{array}{l} \sqrt{\cos^* \theta_{1,1} \cos\theta_{1,2}} \sin\theta_{1,1} \sin\theta_{1,2} \\ \times \left[ \begin{array}{l} +2 J_\ell(\Pi_1) J_\ell(\Pi_2) \left( t_{s,1}^* + t_{p,1}^* \cos^* \theta_{1,1} \right) \left( t_{p,2} + t_{s,2} \cos\theta_{1,2} \right) \\ - \begin{bmatrix} +J_{\ell+2}(\Pi_1) J_{\ell+2}(\Pi_2) \\ +J_{\ell-2}(\Pi_1) J_{\ell-2}(\Pi_2) \end{bmatrix} \left( t_{s,1}^* - t_{p,1}^* \cos^* \theta_{1,1} \right) \left( t_{p,2} - t_{s,2} \cos\theta_{1,2} \right) \end{array} \right] \\ \times \exp\left\{ -i \frac{\pi}{\lambda d} \left[ \rho_{1,1}^2 - \rho_{1,2}^2 \right] \right\} e^{-ikz\left( \cos^* \theta_{1,1} - \cos\theta_{1,2} \right)} \end{array} \right\} d\theta_{1,1} d\theta_{1,2} \text{ ,}$$

(S57)

respectively. The three components of SAM density are

$$S_r=\frac{\varepsilon}{4\omega}\frac{\pi^2|A|^2}{2}\int_0^{\theta_{\max}}\int_0^{\theta_{\max}}\mathrm{Re}\left\{\begin{array}{l}\sqrt{\cos^*\theta_{1,1}\cos\theta_{1,2}}\sin\theta_{1,1}\sin\theta_{1,2}\\ \times\left[\begin{array}{l}-i\sin\theta_{1,1}\begin{bmatrix}+J_\ell(\Pi_2)J_{\ell+1}(\Pi_1)\\+J_\ell(\Pi_2)J_{\ell-1}(\Pi_1)\end{bmatrix}\begin{pmatrix}+t_{p,1}^*\left(t_{s,2}+t_{p,2}\cos\theta_{1,2}\right)\\+t_{s,1}^*\left(t_{p,2}+t_{s,2}\cos\theta_{1,2}\right)\end{pmatrix}\\ +i\sin\theta_{1,1}\begin{bmatrix}+J_{\ell+1}(\Pi_1)J_{\ell+2}(\Pi_2)\\+J_{\ell-1}(\Pi_1)J_{\ell-2}(\Pi_2)\end{bmatrix}\begin{pmatrix}+t_{p,1}^*\left(t_{s,2}-t_{p,2}\cos\theta_{1,2}\right)\\+t_{s,1}^*\left(t_{p,2}-t_{s,2}\cos\theta_{1,2}\right)\end{pmatrix}\\ +i\sin\theta_{1,2}\begin{bmatrix}+J_\ell(\Pi_1)J_{\ell+1}(\Pi_2)\\+J_\ell(\Pi_1)J_{\ell-1}(\Pi_2)\end{bmatrix}\begin{pmatrix}+t_{p,2}\left(t_{s,1}^*+t_{p,1}^*\cos^*\theta_{1,1}\right)\\+t_{s,2}\left(t_{p,1}^*+t_{s,1}^*\cos^*\theta_{1,1}\right)\end{pmatrix}\\ -i\sin\theta_{1,2}\begin{bmatrix}+J_{\ell+1}(\Pi_2)J_{\ell+2}(\Pi_1)\\+J_{\ell-1}(\Pi_2)J_{\ell-2}(\Pi_1)\end{bmatrix}\begin{pmatrix}+t_{p,2}\left(t_{s,1}^*-t_{p,1}^*\cos^*\theta_{1,1}\right)\\+t_{s,2}\left(t_{p,1}^*-t_{s,1}^*\cos^*\theta_{1,1}\right)\end{pmatrix}\end{array}\right]\\ \times\exp\left\{-i\frac{\pi}{\lambda d}\left[\rho_{1,1}^2-\rho_{1,2}^2\right]\right\}e^{-ikz\left(\cos^*\theta_{1,1}-\cos\theta_{1,2}\right)}\end{array}\right\}d\theta_{1,1}d\theta_{1,2}\,,$$

(S58)

$$S_\varphi=\frac{\varepsilon}{4\omega}\frac{\pi^2|A|^2}{2}\int_0^{\theta_{\max}}\int_0^{\theta_{\max}}\mathrm{Re}\left\{\begin{array}{l}\sqrt{\cos^*\theta_{1,1}\cos\theta_{1,2}}\sin\theta_{1,1}\sin\theta_{1,2}\\ \times\left[\begin{array}{l}+\sin\theta_{1,1}\begin{bmatrix}+J_\ell(\Pi_2)J_{\ell+1}(\Pi_1)\\-J_\ell(\Pi_2)J_{\ell-1}\left({}^{1}\Pi\right)\end{bmatrix}\begin{pmatrix}+t_{p,1}^*\left(t_{s,2}+t_{p,2}\cos\theta_{1,2}\right)\\+t_{s,1}^*\left(t_{p,2}+t_{s,2}\cos\theta_{1,2}\right)\end{pmatrix}\\ -\sin\theta_{1,1}\begin{bmatrix}-J_{\ell+1}(\Pi_1)J_{\ell+2}(\Pi_2)\\+J_{\ell-1}(\Pi_1)J_{\ell-2}(\Pi_2)\end{bmatrix}\begin{pmatrix}+t_{p,1}^*\left(t_{s,2}-t_{p,2}\cos\theta_{1,2}\right)\\+t_{s,1}^*\left(t_{p,2}-t_{s,2}\cos\theta_{1,2}\right)\end{pmatrix}\\ +\sin\theta_{1,2}\begin{bmatrix}+J_\ell(\Pi_1)J_{\ell+1}(\Pi_2)\\-J_\ell(\Pi_1)J_{\ell-1}(\Pi_2)\end{bmatrix}\begin{pmatrix}+t_{p,2}\left(t_{s,1}^*+t_{p,1}^*\cos^*\theta_{1,1}\right)\\+t_{s,2}\left(t_{p,1}^*+t_{s,1}^*\cos^*\theta_{1,1}\right)\end{pmatrix}\\ -\sin\theta_{1,2}\begin{bmatrix}-J_{\ell+1}(\Pi_2)J_{\ell+2}(\Pi_1)\\+J_{\ell-1}(\Pi_2)J_{\ell-2}(\Pi_1)\end{bmatrix}\begin{pmatrix}+t_{p,2}\left(t_{s,1}^*-t_{p,1}^*\cos^*\theta_{1,1}\right)\\+t_{s,2}\left(t_{p,1}^*-t_{s,1}^*\cos^*\theta_{1,1}\right)\end{pmatrix}\end{array}\right]\\ \times\exp\left\{-i\frac{\pi}{\lambda d}\left[\rho_{1,1}^2-\rho_{1,2}^2\right]\right\}e^{-ikz\left(\cos^*\theta_{1,1}-\cos\theta_{1,2}\right)}\end{array}\right\}d\theta_{1,1}d\theta_{1,2}\,,$$

(S59)

$$S_z=\frac{\varepsilon}{2\omega}\frac{\pi^2|A|^2}{4}\int_0^{\theta_{\max}}\int_0^{\theta_{\max}}\mathrm{Re}\left\{\begin{array}{l}\sqrt{\cos^*\theta_{1,1}\cos\theta_{1,2}}\begin{bmatrix}-J_{\ell+2}(\Pi_2)J_{\ell+2}(\Pi_1)\\+J_{\ell-2}(\Pi_2)J_{\ell-2}(\Pi_1)\end{bmatrix}\\ \times\begin{pmatrix}+\left(t_{s,1}^*-t_{p,1}^*\cos^*\theta_{1,1}\right)\left(t_{s,2}-t_{p,2}\cos\theta_{1,2}\right)\\+\left(t_{p,1}^*-t_{s,1}^*\cos^*\theta_{1,1}\right)\left(t_{p,2}-t_{s,2}\cos\theta_{1,2}\right)\end{pmatrix}\\ \times\exp\left\{-i\frac{\pi}{\lambda d}\left[\rho_{1,1}^2-\rho_{1,2}^2\right]\right\}e^{-ikz\left(\cos^*\theta_{1,1}-\cos\theta_{1,2}\right)}\sin\theta_{1,1}\sin\theta_{1,2}\end{array}\right\}d\theta_{1,1}d\theta_{1,2}\,,\quad\text{(S60)}$$

respectively. The distributions of the KM densities and SAM densities at different NA in the focusing system are shown in **FIG. S4**. From the Figures, it can be found that, apart from the axial KM density (**FIG. S4**(a), **S4**(e), **S4**(i)), the azimuthal KM density (**FIG. S4**(b), **S4**(f), **S4**(j)) originated from the orbital angular momentum property of vortex light field is present. Moreover, the axial (**FIG. S4**(c), **S4**(g), **S4**(k)) and the azimuthal SAM densities (**FIG. S4**(d), **S4**(h), **S4**(l)) are present simultaneously. This is different from the result given in Reference [39], which declares that the unpolarized light does not possess the SAM component along the propagating direction (axial direction).

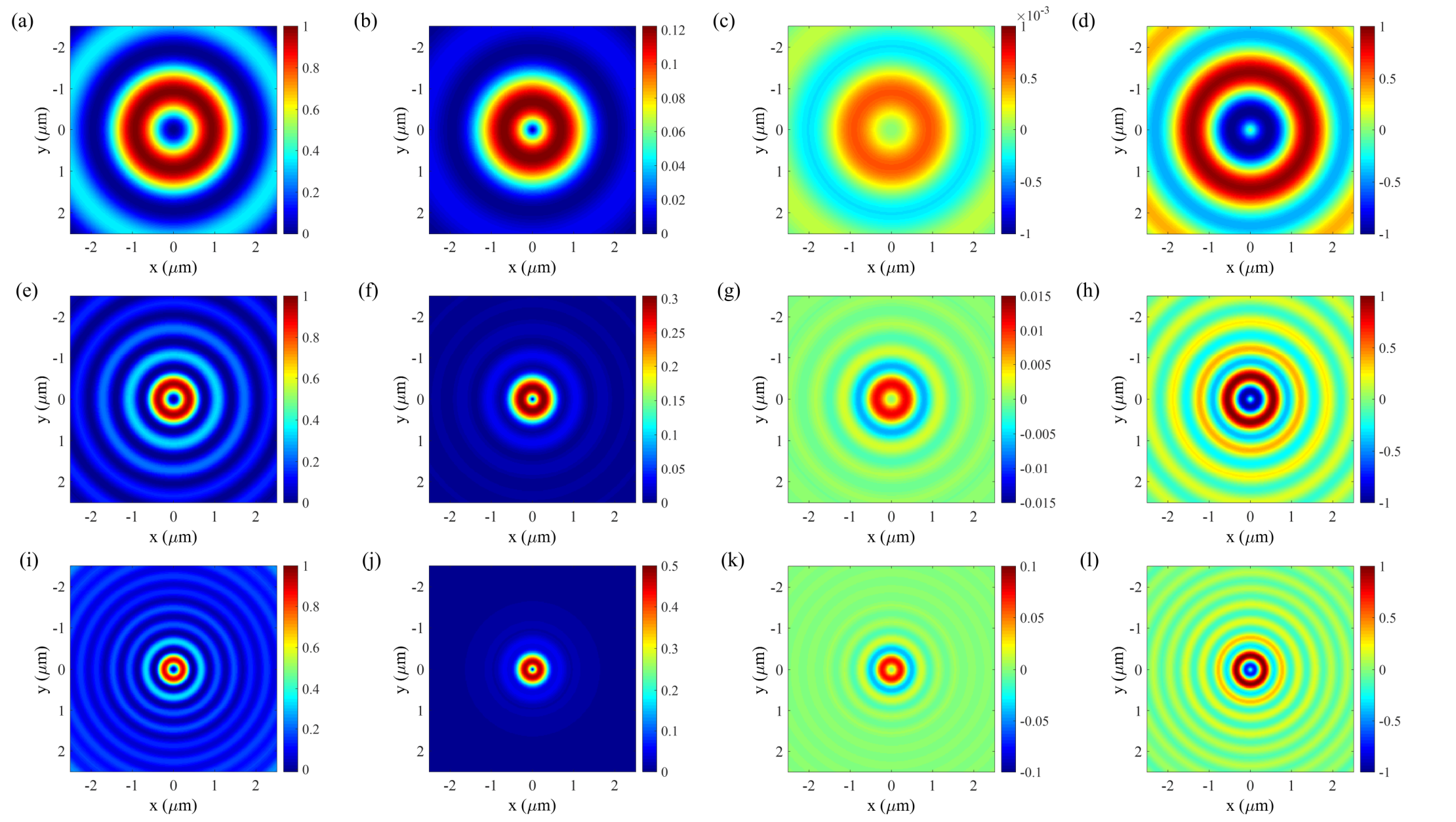

**FIG. 4.** The KM and SAM properties of the focused unpolarized vortex light fields. When NA = 0.2, (a) the axial, (b) the azimuthal KM densities and (c) the axial, (d) the azimuthal SAM densities in the focal plane are shown. (e-h) Same with (a-d) but with NA = 0.5. (i-l) Same with (a-d) but with NA = 0.8. The topological charge of incident vortex light is $\ell = 1$.

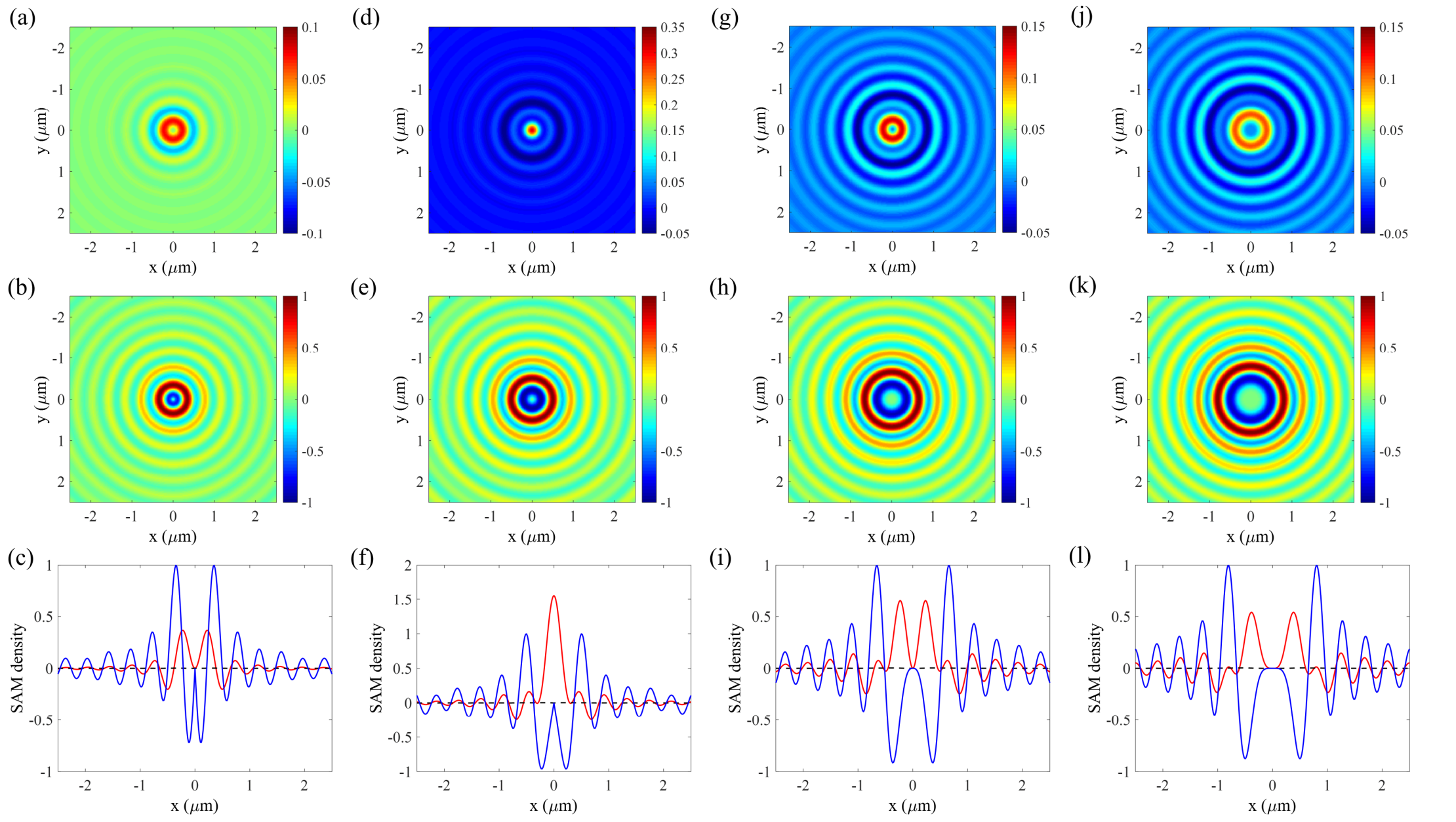

**FIG. 5.** The SAM properties of the focused unpolarized vortex light fields. When the topological charge of incident vortex light is $\ell = 1$, (a) the axial, (b) the azimuthal SAM densities and (c) the one-dimensional (1D) contour of these SAM densities (red line for axial SAM and blue line for azimuthal SAM) in the $x$-direction are shown. (d-f) Same with (a-c) but with $\ell = 2$. (g-i) Same

with (a-c) but with $\ell$ = 3. (j-l) Same with (a-c) but with $\ell$ = 4. Interestingly, the axial SAM density possesses a Gaussian distribution when topological charge of incident vortex light is $\ell$ = 2, and it is hollow otherwise. Moreover, from the 1D contour of the SAM densities, one can find that the spin vectors construct the domain wall-like spin textures, and the skyrmion number of these spin textures is zero. These results are totally different from those produced by the focused polarized vortex light field. NA = 0.8 is used here.

In these instances, in line with Expression (S50), the SAM density is also locked with the KM density except for an additional coupling term $\mathbf{S}_c$. However, comparing to the aforementioned situation, the azimuthal KM density and the axial SAM density arise from the orbital angular momentum the orbital angular momentum property of vortex light field and their magnitudes enlarge as the NA increases. From the spin-momentum locking relationship given by Equation (S50), the azimuthal SAM density (**FIG. S4**(d), **S4**(h), **S4**(l)) is derivated from the radial gradient of the axial KM density (**FIG. S4**(a), **S4**(e), **S4**(i)) and the axial SAM density (**FIG. S4**(c), **S4**(g), **S4**(k)) is related to the radial gradient of the azimuthal KM density (**FIG. S4**(b), **S4**(f), **S4**(j)).

Then, we also investigate the topological defects formed by the spin vectors of these focused unpolarized vortex light fields. The distributions of the SAM densities for the focused unpolarized vortex light fields with vortex topological charge $\ell$ = 1, 2, 3, 4 respectively are shown in **FIG. S5**. It can be observed that there are the axial SAM densities (**FIG. S5**(a), **S5**(d), **S5**(g), **S5**(j)) and the azimuthal SAM densities (**FIG. S5**(b), **S5**(e), **S5**(h), **S5**(k)), which is a manifestation of Bloch-type topological defects. However, by carefully examining the one-dimensional (1D) contour of these SAM densities (**FIG. S5**(c), **S5**(f), **S5**(i), **S5**(l)), one can realize that the domain wall-like spin textures are formed by these focused unpolarized vortex light fields, and the skyrmion numbers of these spin textures are equal to zero. This is the first difference between the topological defects constructed by a focused polarized vortex light and a focused unpolarized light.

Moreover, we notice that the axial SAM density possesses a Gaussian distribution (0-order Bessel function) when topological charge of incident vortex light is $\ell$ = 2, and it is hollow otherwise ($\ell$ > 0). This is the second difference between the topological defects constructed by a focused polarized vortex light and a focused unpolarized light. Because for the photonic spin skyrmions constructed by a focused polarized vortex light, the axial SAM density should possess a Gaussian distribution (0-order Bessel function) when topological charge of incident vortex light is $\ell$ = 1, and it is hollow otherwise ($\ell$ > 0).

Particularly, from Equation (S50), the coupling term $\mathbf{S}_c$ can be cancelled by a polarizer, which only allows the $p$-polarization (transverse magnetic polarized modes) or the s-polarization (transverse electric polarized modes) pass through the polarizer. Here, we consider the surface plasmon (SP) mode excited by this focused unpolarized vortex light at the metal/air interface. By substituting $\cos\theta_{1,1} = ik_{z,1}/k$, $\cos\theta_{1,2} = ik_{z,2}/k$, $\sin\theta_{1,1} = k_{\perp,1}/k$, $\sin\theta_{1,2} = k_{\perp,2}/k$ and $\theta_{1,1} = \theta_{1,2}$ into the Expressions (S55-S60), we can obtain the KM density and SAM density of SP modes excited by this focused unpolarized vortex light as

$$P_r = 0, \tag{S61}$$

$$P_\varphi = \frac{\pi^2\eta\left|t_p A\right|^2}{4c^2}\frac{k_{z,1}}{k}\frac{k_{\perp,1}^3}{k^3}\left[\frac{2(\ell+1)}{\Pi}J_{\ell+1}(\Pi)J_{\ell+1}(\Pi)+\frac{2(\ell-1)}{\Pi}J_{\ell-1}(\Pi)J_{\ell-1}(\Pi)\right]e^{-2k_{z,1}z}, \tag{S62}$$

$$P_z = 0, \tag{S63}$$

$$S_r = \frac{\pi^2\varepsilon\left|t_p A\right|^2}{4\omega}\frac{k_{z,1}^2}{k^2}\frac{k_{\perp,1}^2}{k^3}\left[\frac{2(\ell+1)}{\Pi}J_{\ell+1}(\Pi)J_{\ell+1}(\Pi)+\frac{2(\ell-1)}{\Pi}J_{\ell-1}(\Pi)J_{\ell-1}(\Pi)\right]e^{-2k_{z,1}z}, \tag{S64}$$

$$S_\varphi = 0, \tag{S65}$$

$$S_z = \frac{\pi^2\varepsilon\left|t_p A\right|^2}{8\omega}\frac{k_{z,1}}{k}\frac{k_{\perp,1}^4}{k^4}\left[J_{\ell-2}(\Pi)J_{\ell-2}(\Pi)-J_{\ell+2}(\Pi)J_{\ell+2}(\Pi)\right]e^{-2k_{z,1}z}. \tag{S66}$$

Here, we assume that only the single wave vector component satisfying the wave vector matching condition [53] can pass through the metal and form the SP modes at the metal/air interface. By simple derivations, we can obtain that the KM density and SAM density of SP modes excited by this focused unpolarized vortex light satisfy the intrinsic spin-momentum locking relation, i.e.,

$$\mathbf{S} = \frac{1}{2k^2}\nabla\times\mathbf{P}. \tag{S67}$$

Since the KM density and SAM density of the focused unpolarized light also satisfy the intrinsic spin-momentum locking relation, the topological defects can be also constructed by the focused unpolarized light.

The SAM densities of the focused unpolarized light at metal interface with vortex topological charge $\ell$ = 1, 2, 3, 4 respectively are shown in **FIG. S6**. It can be observed that there are the radial SAM densities (**FIG. S6**(a), **S6**(d), **S6**(g), **S6**(j)) and the normal components of SAM densities (**FIG. S6**(b), **S6**(e), **S6**(h), **S6**(k)), which is a manifestation of Néel-type topological defects. It can be observed that the normal component of SAM density also possesses a Gaussian distribution when topological charge of incident vortex light is $\ell$ = 2, and it is hollow otherwise. This is the first difference between the topological defects constructed by the SP modes excited by a focused polarized vortex light and a focused unpolarized light.

Moreover, by carefully examining the one-dimensional (1D) contour of these SAM densities (**FIG. S6**(c), **S6**(f), **S6**(i), **S6**(l)), we notice that the skyrmion number of these spin textures is not always 1, especially at the center for $\ell > 1$ (as shown in the green circles). The skyrmion number of these spin textures is 1 only after the first zero-point of radial SAM density (period > $\ell-1$). This is the second difference between the topological defects constructed by the SP modes excited by a focused polarized vortex light and a focused unpolarized light.

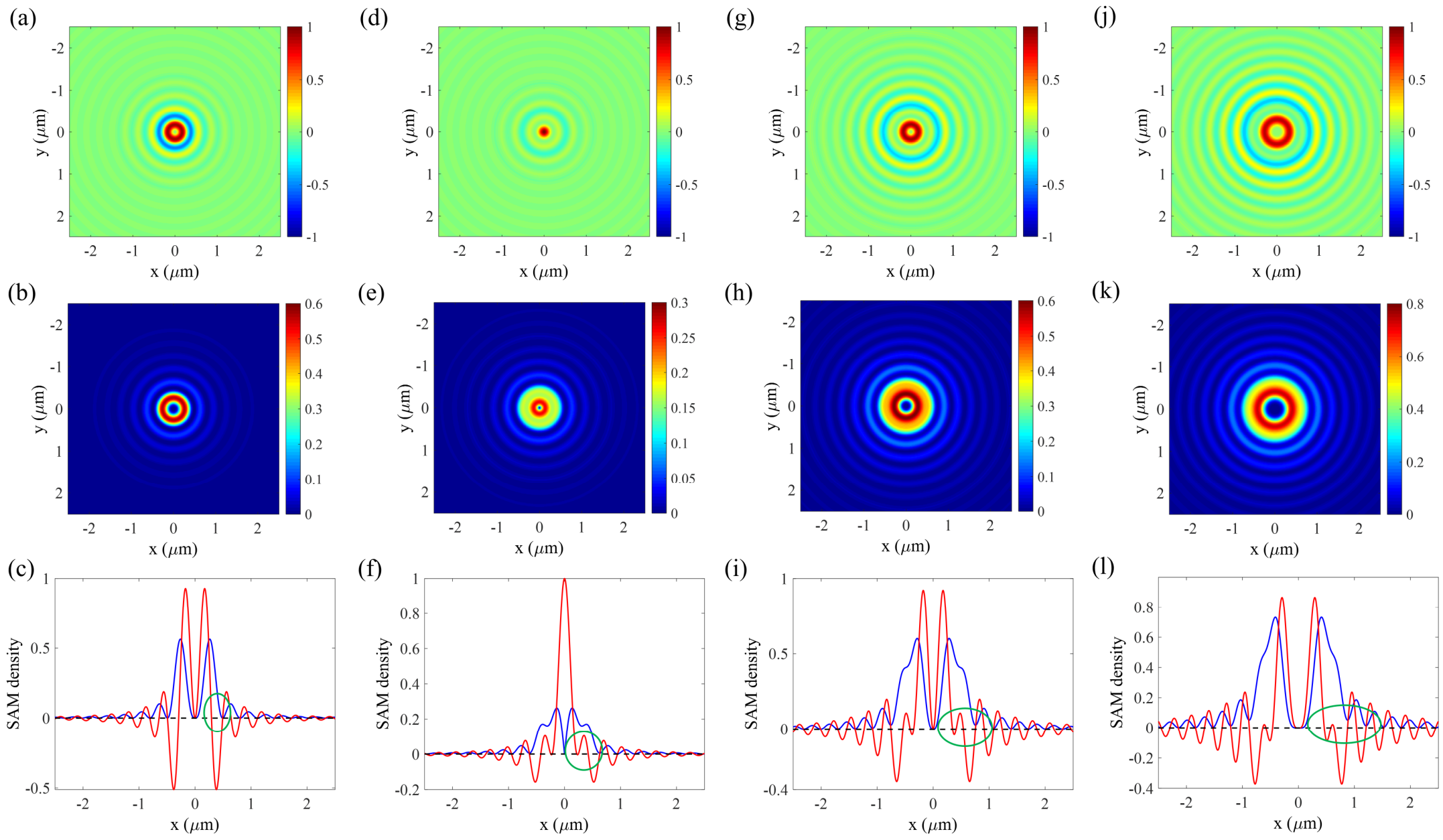

**FIG. 6.** The SAM properties of the focused unpolarized vortex light fields at the metal/air interface. When the topological charge of incident vortex light is $\ell = 1$, (a) the normal, (b) the radial components of SAM densities and (c) the 1D contour of the SAM components (red line for normal SAM and blue line for radial SAM) in the $x$-direction are shown. (d-f) Same with (a-c) but with $\ell = 2$. (g-i) Same with (a-c) but with $\ell = 3$. (j-l) Same with (a-c) but with $\ell = 4$. The normal component of SAM density possesses a Gaussian distribution when topological charge of incident vortex light is $\ell = 2$, and it is hollow otherwise. Moreover, from the 1D contour of the SAM densities, one can find that the spin vectors construct the Néel-type skyrmion-like spin textures. However, we notice that the skyrmion number of these spin textures is 1 only after the first zero-point of radial SAM density. These results are totally different from those produced by the focused polarized vortex light field. $k_\perp = 1.05k_0$ is used here.

## V. Experimental setup and measurement of SAM density of the focused unpolarized vortex light at the metal/air interface

The experimental setup of the mapping the out-of-plane component of SAM densities of the focused unpolarized light with various vortex topological charges is given in **FIG. S7**.

The incident light of 632.8nm-wavelength passes through a linear polarizer (LP), a half-wave plane (HWP) and a spatial light modulator (SLM) to generate a linear polarized light carrying the desired vortex phase. Then, this light passes a HWP and 1-order vortex wave plate to generate a radially polarized light carrying the desired vortex phase. Subsequently, the combination of two liquid crystal (LC) plates and a quarter-wave plate (QWP) is inserted to make sure that the polarization of light can be transformed into the arbitrary polarized state in the surface of Poincaré sphere. For example, the Jones matrix of linear polarized light is

$$\mathbf{E}_0 = \begin{bmatrix} 1 \\ 0 \end{bmatrix}. \tag{S68}$$

The intersection angle between the fast axis of the LC1 and the polarization direction of the linear polarized light is 45°, and thus the Jones matrix of LC1 is

$$\mathbf{J}_{\mathrm{LC1}}=\cos\frac{\delta_1}{2}\begin{bmatrix}1 & 0\\ 0 & 1\end{bmatrix}-i\sin\frac{\delta_1}{2}\begin{bmatrix}0 & 1\\ 1 & 0\end{bmatrix}. \quad \text{(S69)}$$

Here, $\delta_1$ represents the phase difference between the fast axis and the slow axis of LC1. $i$ is the imaginary unit.

For the LC2, the intersection angle between the fast axis and the polarization direction of the linear polarized light is 0º, and thus its Jones matrix is

$$\mathbf{J}_{\mathrm{LC2}}=\cos\frac{\delta_2}{2}\begin{bmatrix}1 & 0\\ 0 & 1\end{bmatrix}-i\sin\frac{\delta_2}{2}\begin{bmatrix}1 & 0\\ 0 & -1\end{bmatrix}. \quad \text{(S70)}$$

Here, $\delta_2$ represents the phase difference between the fast axis and the slow axis of LC2.

In addition, the the intersection angle between the fast axis of QWP and the polarization direction of the linear polarized light is −45º, and thus its Jones matrix is

$$\mathbf{M}=\frac{1}{\sqrt{2}}\begin{bmatrix}1 & i\\ i & 1\end{bmatrix}. \quad \text{(S71)}$$

From the expressions, we can calculate the Jones matrix of the linear polarized light passing through the combination of LC1, LC2 and QWP as

$$\mathbf{E}=\mathbf{M}\mathbf{J}_{\mathrm{LC2}}\mathbf{J}_{\mathrm{LC1}}\mathbf{E}_0=\frac{1}{\sqrt{2}}\cos\left(\frac{\delta_1}{2}\right)e^{-i\frac{\delta_2}{2}}\begin{bmatrix}1\\ i\end{bmatrix}+\frac{1}{\sqrt{2}}\sin\left(\frac{\delta_1}{2}\right)e^{+i\frac{\delta_2}{2}}\begin{bmatrix}1\\ -i\end{bmatrix}. \quad \text{(S72)}$$

This expression is consistent with the polarization states in a Poincaré sphere [50]. To generate the polarization state covering any point on the surface of the Poincare sphere, we can control the phase delay of the LC device by the random voltage generated by the acquisition card.

Finally, this Poincaré light beam is tightly focused by an oil immersion objective lens (Olympus, NA=1.49, 100×) to a PS nanoparticle with diameter 201nm sitting on a gold film (thickness: 45nm). The position of the PS particle can be precisely controlled by a piezo-stage (Physik Instrumente, P-545). The near-field signal that is scattered by the PS particle to an objective lens (Olympus, NA = 0.7, 60×) was split and then analyzed using a combination of a QWP and a LP to extract the individual circular polarization component ($I_{\mathrm{LCP}}$: left-handed circularly polarized component and $I_{\mathrm{RCP}}$: right-handed circularly polarized component) of the signal. Through the dipole theory [51], the horizontal components of electric field at each time can be reconstructed. These components were then directed to two photomultiplier tubes (PMTs) to measure the intensity information of the two signals. This then enables characterization of the out-of-plane SAM component (i.e., along the optical axis) of the focused beams by the relation [17]:

$$S_x=\frac{\varepsilon}{4\omega}\frac{k_{\perp,1}^2}{k_{z,1}^2}\left(I_{\mathrm{RCP}}-I_{\mathrm{LCP}}\right). \quad \text{(S73)}$$

In this experiment, the experimental data of each pixel is the average of the collected results under the incident of multiple Poincaré polarized states. Here, we set the number of Poincaré polarized states as 50, and the pixel of image are 61*61 (the number of random voltages to be applied in the experiment is 50*61*61). The time interval between applying random voltages is 2ms, and we takes about 12min to obtain an image of SAM density.

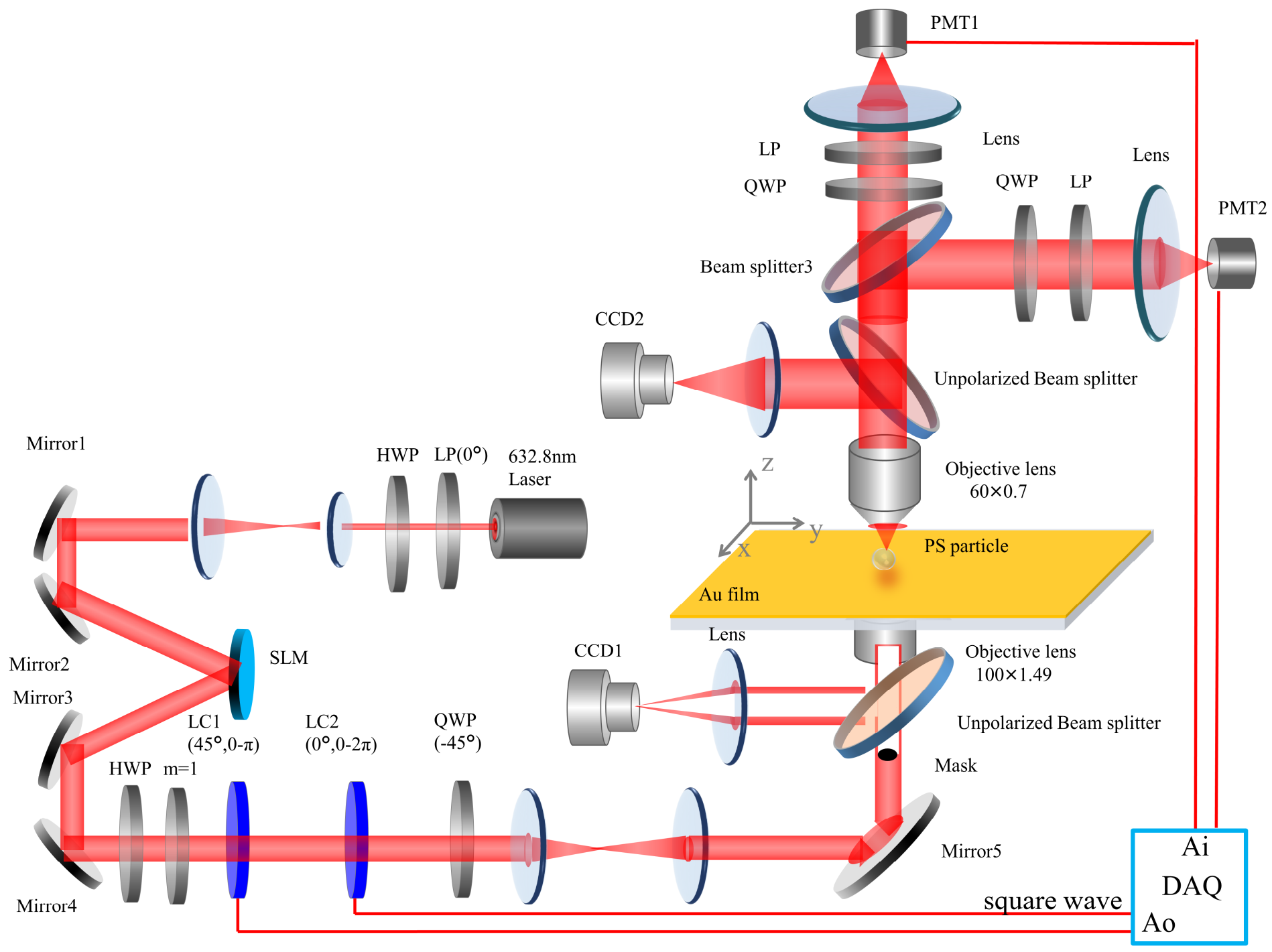

**FIG. S7.** The schematic diagram for mapping the normal component of SAM densities for the focused unpolarized vortex light fields at the metal/air interface.

The theoretical and reconstructed results for the focused left-handed circularly polarized light and right-handed circularly polarized plane wave can be found in **FIG. S8**. These experimental results match well with the theoretical results, which demonstrate the validity of the proposed spin-momentum locking theory.

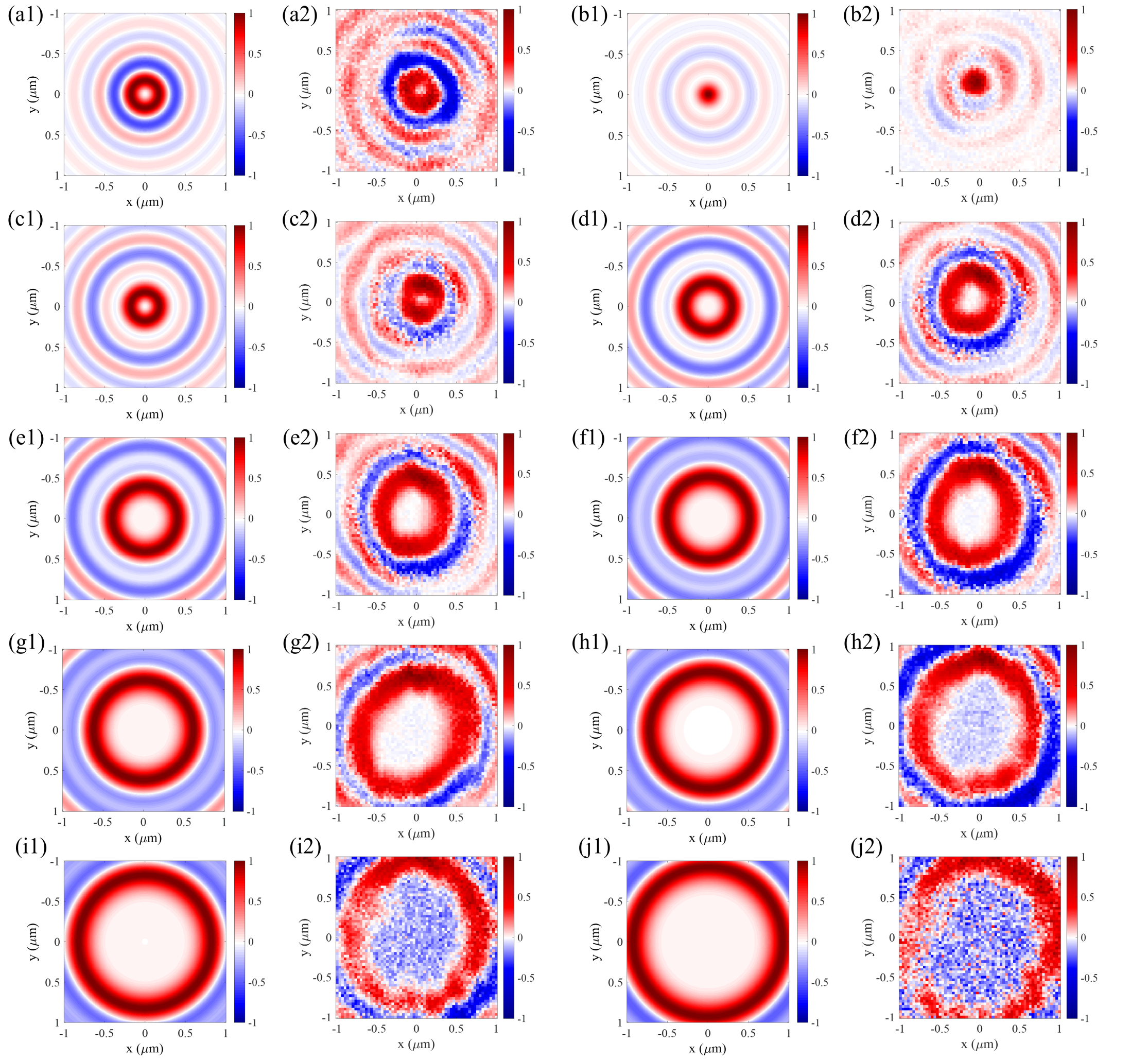

**FIG. S8.** (a1) The theoretical $S_z$ and (a2) the experimental $S_z$ with $\ell = 1$; (b1-b2) Same as (a1-a2) but with $\ell = 2$; (c1-c2) Same as (a1-a2) but with $\ell = 3$; (d1-d2) Same as (a1-a2) but with $\ell = 4$; (e1-e2) Same as (a1-a2) but with $\ell = 5$; (f1-f2) Same as (a1-a2) but with $\ell = 6$; (g1-g2) Same as (a1-a2) but with $\ell = 7$; (h1-h2) Same as (a1-a2) but with $\ell = 8$; (i1-i2) Same as (a1-a2) but with $\ell = 9$; (j1-j2) Same as (a1-a2) but with $\ell = 10$. The theoretical results match well with the experimental results, which demonstrate the validity of the theory. Here, the region of image is 2μm×2μm uniformly, and the grid size is 40nm, and the wavelength is 632.8nm.